Review article

The origin of ultrasensitive SERS sensing beyond plasmonics

Leilei Lan, Yimeng Gao, Xingce Fan, Mingze Li, Qi Hao, Teng Qiu†
*School of Physics, Southeast University, Nanjing 211189, China*
*Corresponding author. E-mail: †tqiu@seu.edu.cn*

# The origin of ultrasensitive SERS sensing beyond plasmonics

Leilei Lan, Yimeng Gao, Xingce Fan, Mingze Li, Qi Hao and Teng Qiu†

School of Physics, Southeast University, Nanjing, 211189, P. R. China

Corresponding author. E-mail: †tqiu@seu.edu.cn

## Abstract

Plasmon-free surface-enhanced Raman scattering (SERS) substrates have attracted tremendous attention for their abundant sources, excellent chemical stability, superior biocompatibility, good signal uniformity, and unique selectivity to target molecules. Recently, researchers have made great progress in fabricating novel plasmon-free SERS substrates and exploring new enhancement strategies to improve their sensitivity. This review summarizes the recent developments of plasmon-free SERS substrates and specially focuses on the enhancement mechanisms and strategies. Furthermore, the promising applications of plasmon-free SERS substrates in biomedical diagnosis, metal ions and organic pollutants sensing, chemical and biochemical reactions monitoring, and photoelectric characterization are introduced. Finally, current challenges and future research opportunities in plasmon-free SERS substrates are briefly discussed.

**Contents**

## 1 Introduction

As pioneers of surface-enhanced Raman scattering (SERS), *Fleischmann et al.* discovered the enhanced Raman scattering effect in 1974 [1], who initially attributed the signals increase to a higher number of adsorbed molecules with the increased surface area during measurements of the Raman scattering of pyridine on an electrochemically roughened silver electrode. This phenomenon was identified in 1977 by *Jeanmaire and Van Duyne* and *Albrecht and Creighton* independently [2, 3]. They both recognized that the tremendously strong surface Raman signals could not be explained by the increased surface area curtly and demonstrated the increased signals derived from a true enhancement of the Raman scattering efficiency itself, which created an exciting area of Raman spectroscopy, namely SERS. As one of the most powerful analytical tools, SERS overcomes the inherent shortcoming of weak signals in traditional Raman scattering with the magnification more than one million times. Besides, SERS requires less for the sample pretreatment and is more suitable for diverse analytical systems compared with mass spectrometry, fluorescence spectroscopy and other spectroscopy techniques. Over the past decades, tens of thousands research papers about SERS have been published, with SERS widely applied in various fields such as surface science, photonics, biomedicine, and trace analysis [4-10]. For example, the combination of scanning probe microscopy with SERS has led to a new field, tip-enhanced Raman scattering (TERS), which provides abundant chemical signatures at high detection sensitivity up to single molecule and high spatial resolution up to (sub)nanometer [11, 12]. By the shell-isolated nanoparticle-enhanced Raman spectroscopy (SHINERS) technique, *Li et al.* studied the *in-situ* electrochemical behavior of different molecules and catalytic reactions [13]. In addition, *Gu et al.* demonstrated a physical unclonable functions label fabricated by drop-casting aqueous gap-enhanced Raman tags, which provides a potential platform to realize unbreakable anticounterfeiting [14]. However, high SERS activity almost only appears in the plasmonic noble-metal materials (mainly Au and Ag), and largely relies on the density of electromagnetic ‘hot spots’ on the rough surface, which extremely limits the selection of materials [15]. Besides, these designed structures are particularly vulnerable to the environmental interference, indicating inferior stability. Apart from that, there are some other drawbacks including high cost and poor biocompatibility [16].

Compared with traditional plasmonic-based SERS, plasmon-free SERS, as a new frontier, possesses some special features. Firstly, plasmon-free SERS materials have abundant sources, including transition metal oxides [16-22], transition metal dichalcogenides (TMDs) [23-30], perovskites [31, 32], graphene [33-38], boron nitride [39-41], metal organic frameworks (MOFs) [42-45], organic semiconductors [46, 47], black phosphorus [48-51], and transition metal carbides and nitrides (MXenes) [52-58], which enable multiple choices to realize Raman enhancement [59]. Secondly, plasmon-free SERS materials do not require well-designed 'hot spots', thus showing good signal uniformity and excellent stability, which could adapt to different detection environments. Moreover, multifunctional plasmon-free SERS materials are widely used in photoelectric devices, energy conversion, energy storage, catalysis and so on, which further expand the application scope of SERS [59, 60]. In addition, most of plasmon-free SERS materials are of low price, superior biocompatibility and recyclability. Whereas, the inferior detection sensitivity impedes the practical applications of plasmon-free SERS. Fortunately, some recent studies have made remarkable progress in obtaining ultrasensitive plasmon-free SERS substrates [47, 59, 61-63]. For instance, Ling *et al.* reported that graphene can be used as SERS substrates with a limit of detection (LOD) of $10^{-8}$ M [33]. *Demirel et al.* demonstrated that organic semiconductors could realize noble metal-comparable SERS enhancement and sensitivity by molecular engineering [47]. *Seo et al.* obtained the ultrasensitive plasmon-free SERS substrate with femtomolar detection limit on a two-dimensional (2D) van der Waals heterostructure [61]. As we know, the huge enhancement effect in SERS originates from the interaction between the SERS-active materials and probe molecules, generally including electromagnetic mechanism (EM) and chemical mechanism (CM) [6, 62]. In comparison to the EM-dominated plasmonic materials, plasmon-free materials own diverse surface physical and chemical properties, leading to more complicated enhancement mechanism. In the past for a long time, it was generally recognized that CM associated with charge transfer (CT) played a major role in plasmon-free SERS, based on which the enhancement factor (EF) is usually lower than $10^3$ [62]. Nevertheless, according to the recent reports, there emerged more and more novel plasmon-free SERS substrates and the record of EF was continually broken, making the CT insufficient to explain the mechanism completely. As a result, we have to rethink the intrinsic nature of plasmon-free SERS.

Recent years, a series of new enhancement strategies including defect engineering [17-19], amorphization treatment [64-66], phase engineering [25, 27], heterojunction engineering [61, 67, 68], facet engineering [69, 70], molecular engineering [46, 47], and so on have been used to improve the sensitivity of plasmon-free SERS materials, providing cases for understanding its enhancement mechanisms. These ultrasensitive plasmon-free SERS substrates expand the applications of plasmon-free SERS into various fields, ranging from bioanalysis to photoelectric characterization.

This review focuses on the background, enhancement mechanisms, enhancement strategies, and applications of ultrasensitive SERS sensing beyond plasmonics. At first, we introduce the background of SERS and the advantages of plasmon-free SERS. Afterwards, we discuss the enhancement mechanisms of plasmon-free SERS materials, followed by some strategies for improving their SERS sensitivity. Moreover, the applications of plasmon-free SERS materials in the areas of biomedical diagnosis, metal ions and organic pollutants sensing, chemical and biochemical reactions monitoring, and photoelectric characterization are summarized. Finally, future trend of plasmon-free SERS is briefly discussed.

## 2 The mechanisms of plasmon-free SERS

In recent years, as the plasmon-free SERS flourishes, the researches on its enhancement mechanisms have made remarkable progress. Until now, plenty of studies indicate that plasmon-free SERS strongly rely on material size, surface defect, sample morphology, crystallinity, and crystal orientation [64-72]. Some associated phenomena including Mie resonance, CT resonance, exciton resonance and molecular resonance may play important roles solely or synergistically.

### 2.1 Mie resonance

Mie resonance, known as morphology-dependent resonance, can give rise to strong local electromagnetic field and promote the interaction between light and matter. Compared with localized surface plasmon resonance (LSPR) in metallic particles, Mie resonance typically exists in dielectric particles [73, 74]. The Mie scattering can cause Raman enhancement when the particle size is comparable to the wavelength of incident light. This situation can be identified by an important size parameter $x$ (,*ie.* $2\pi r/\lambda$), where r is the radius of the spherical particle, and $\lambda$ is the wavenumber of the incident light. The Mie regime is defined for $0.1 < x < 100$. The scattering efficiency is the ratio of the power of the scattered light to that of the incident light [75]. Particularly, near dielectric sphere surfaces, the amplitude of the near-field scattering can be much larger than that of the far-field scattering, leading to a local electromagnetic field enhancement. According to Mie theory, the near-field scattering efficiency $Q_{NF}$ is defined to evaluate the ability of a spherical particle to convert incident electric-field intensity into electric near-field intensity, which can be presented as follow [75]:

$$Q_{NF} = 2\sum_{n=1}^{\infty}\left\{|a_n|^2\left[(n+1)\left|h_{n-1}^{(2)}(x)\right|^2 + n\left|h_{n+1}^{(2)}(x)\right|^2\right] + (2n+1)|b_n|^2\left|h_n^{(2)}(x)\right|^2\right\} \quad (1)$$

where $h_n^{(2)}$ is the second kind of Hankel function, $a_n$ and $b_n$ are the scattering coefficients, which can be expressed by using the complex Riccati-Bessel functions $\psi$ and $\zeta$ [76]:

$$a_n = \frac{\psi_n'(mx)\psi_n(x) - m\psi_n(mx)\psi_n'(x)}{\psi_n'(mx)\zeta_n(x) - m\psi_n(mx)\zeta_n'(x)} \quad (2)$$

$$b_n = \frac{m\psi_n'(mx)\psi_n(x) - \psi_n(mx)\psi_n'(x)}{m\psi_n'(mx)\zeta_n(x) - \psi_n(mx)\zeta_n'(x)} \quad (3)$$

where $m = n + ik$ is the complex refractive index, $n$ is the real refractive index and $k$ is extinction coefficient part of the complex refractive. It should be noted that the complex refractive index is a fundamental physical property of a material, determined by the wavelength of incident light. Consequently, for a specific material, the near-field scattering intensity mainly depends on the size of spherical particle and the wavelength of incident light. Relevant theoretical studies indicate that the EF is approximately proportional to the second power of $Q_{NF}$ [77]. Thus, Mie resonance provides giant Raman enhancement for dielectrics that are not able to generate LSPR in the visible region.

As early as 1988, *Hayashi et al.* observed increased Raman scattering from copper phthalocyanine (CuPc) molecules adsorbed on GaP nanoparticles with different sizes and explained the Raman enhancement in plamon-free materials by Mie scattering theory for the first time [78]. Recently, some groups reported significantly enhanced Raman scattering on plasmon-free materials through morphology-design-induced Mie resonance. Without plasmonic enhancers, *Alessandri et al.* observed remarkable Raman scattering enhancement on $TiO_2$ shell-based spherical resonators, which can be explained by multiple light scattering through the spheres, high refractive index of the shell layer, and related geometrical factors [79]. Further, they also demonstrated that the Mie resonance could be generated in 2μm-size $SiO_2/ZrO_2$ core/shell beads utilized as all-dielectric Raman enhancers [80]. *Rodriguez et al.* reported that non-spherical silicon nanoparticles could generate strong SERS signals, this obvious enhancement stems from huge evanescent electromagnetic fields associated with the Mie resonance [81]. Besides, a remarkable enhancement of Raman sensitivity was obtained on submicron-sized spherical ZnO superstructures by *Ji et al.* (Fig. 1), which was attributed to the synergistic effect of CT in ZnO nanocrystals and Mie resonance of the superstructures [77].

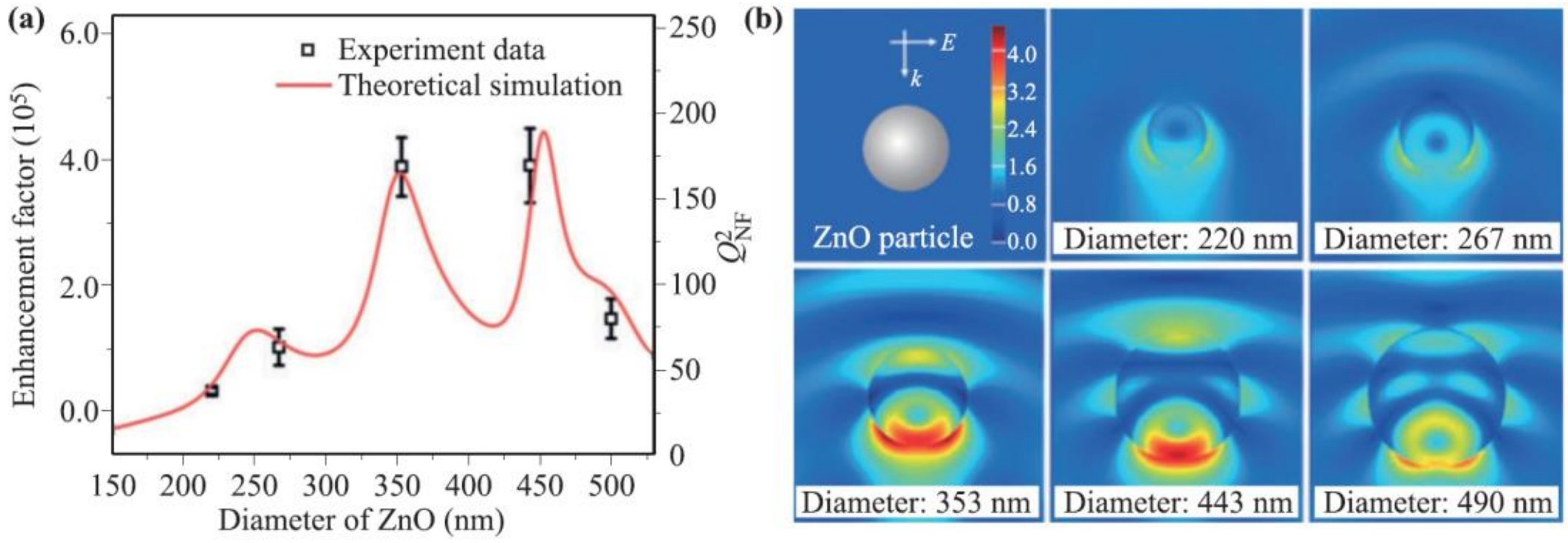

**Fig. 1** (a) The EF and (b) electric near-field distribution of the ZnO particles with different diameters at a 532 nm excitation. (Reprinted and adapted with permission from ref. [77]. Copyright 2019 Wiley-VCH.)

## 2.2 Charge transfer resonance

CT resonance was first used to explain CM enhancement in plasmonic metals as EM mechanism fails to deal with all the SERS phenomena. For example, although CO and $N_2$ have similar Raman cross section, the Raman EF of CO is 200 times higher than that of $N_2$ under identical experimental conditions, which is unreasonable according to EM mechanism [82]. *Jensen et al.* summarized three different situations where the SERS signals are enhanced due to the CM enhancement: (a) enhancement due to ground state chemical interactions between the molecule and nanoparticle which are not associated with the excitation wavelength of the SERS system, (b) resonance Raman enhancement with the excitation wavelength resonant with a molecular transition, and (c) CT resonance Raman enhancement when the excitation wavelength is resonant with the molecule-nanoparticle CT transitions [83]. Among these situations, ground state chemical enhancement is also referred to as the static chemical enhancement or non-resonance chemical enhancement, which is independent of the excitation wavelength because the Raman scattering process itself proceeds in a conventional fashion through virtual energy levels. However, it could also generate $10^0$-$10^2$ Raman signal enhancements by the distortion and modification of the molecular electronic and skeleton structures. In order to explain the effect of resonance effect on Raman spectroscopy, fundamental quantum theorem of the Raman process is considered, in which the Raman process is expressed as a second order perturbation, including twofold interactions between molecules and light. The general expression for Raman polarization tensor $\alpha_{\sigma\rho}$ can be written by the Kramers-Heisenberg-Dirac dispersion formula [84, 85]:

$$(a_{\sigma\rho})_{if} = \sum_{n\neq i,f}\left[\frac{\langle i|M_{\sigma}|n\rangle\langle n|M_{\rho}|f\rangle}{E_n - E_i - \hbar\omega_0 - \mathrm{i}\Gamma_n} + \frac{\langle i|M_{\rho}|n\rangle\langle n|M_{\sigma}|f\rangle}{E_n - E_f + \hbar\omega_0 - \mathrm{i}\Gamma_n}\right] \tag{4}$$

where σ and ρ are the scattered and incident polarization directions, respectively; |i>, |n> and |f> are the wavefunctions of the initial state, the intermediate state, and the final state of the molecular systems, respectively; $E_i$, $E_n$, and $E_f$ are the energies of these states; $M_\sigma$ and $M_\rho$ are the electronic dipole moments in the directions of the

Raman excitation and Raman scattering, respectively; $\hbar\omega_0$ is the energy of the excitation photon; $\Gamma_n$ is the inverse of the dephasing time of the intermediate state. The first term $\frac{\langle i|M_\sigma|n\rangle\langle n|M_\rho|f\rangle}{E_n - E_i - \hbar\omega_0 - \mathrm{i}\Gamma_n}$ can correspond to a resonance because the value of $E_n - E_f - \hbar\omega_0$ can be zero; the second term $\frac{\langle i|M_\rho|n\rangle\langle n|M_\sigma|f\rangle}{E_n - E_f + \hbar\omega_0 - \mathrm{i}\Gamma_n}$ cannot represent a resonance, which is customarily neglected in a Raman process. Thus, the Raman resonance depends on the photon excitations of the molecule-metal system. Three different situations, including off-resonance ($E_n - E_f >> \hbar\omega_0$), pre-resonance ($E_n - E_f \sim \hbar\omega_0$), and rigorous-resonance ($E_n - E_f = \hbar\omega_0$) should be considered in Raman resonance. Generally, $\Gamma_n$ is in the order of 100 $cm^{-1}$, $E_n - E_f - \hbar\omega_0 > 10000$ $cm^{-1}$ for off-resonance Raman scattering, and $E_n - E_f - \hbar\omega_0$ is about 0 $cm^{-1}$ under the pre-resonance and rigorous-resonance. Therefore, the value of $\alpha_{\sigma\rho}$ for pre-resonance and rigorous-resonance is ~$10^2$ times larger than the off-resonance. The Raman intensity is in proportion to the square of the $\alpha_{\sigma\rho}$. So, a resonance Raman intensity is ~$10^4$ times larger than off-resonance Raman intensity [86]. Resonance effect in CM enchantment can be divided into two main types: (a) classical resonance effect of Raman scattering, namely, molecular resonance and (b) the spectral changes caused by the CT processes in the molecule-substrate system, namely, CT resonance. Initially, CT resonance was proposed by *Gersten and Burstein,* respectively [87, 88]. While the direct evidence did not be given until 1981. In this year, *Demuth et al.* found the CT between the Fermi energy in metal and molecular energy levels by observing the excited electronic states in pyridine molecules [89]. Since 1980s, a series of SERS materials such as NiO [90], GaP [78] and $TiO_2$ [91] have been discovered and the CT resonance is therefore extended to the plasmon-free SERS materials. Taking semiconductor as an example, in the semiconductor-molecule system, the CT process would strongly depend on the efficiency of the vibronic coupling between the conduction band (CB) and valence band (VB) of semiconductors with the highest occupied molecular orbital (HOMO) and the lowest unoccupied molecular orbital (LUMO) of the adsorbed molecules. According to the thermodynamically permissible transitions in a semiconductor-molecule system, when the energy levels of semiconductors and molecules are matched well with each other, five possible CT pathways may occur, including molecule HOMO-to-CB, CT complex-to-CB, VB-to-molecule LUMO, surface state-to-molecule LUMO, and CB-to-molecule HOMO (Fig. 2) [62]. It should be noted that CT complex is formed by the strong chemical bonding between the molecule and semiconductor. Recently, many strategies have been reported to effectively increase the CT pathways in the system, leading to strong SERS effect [19, 31, 65, 67]. For example, by altering the wavelength of excited laser (473, 532, and 785 nm), *Yu et al.* found that a CT resonance could occur on the $MAPbCl_3$/4-Mpy interface when the excitation wavelength (532 nm) matches the band-band transition energy (2.32 eV) between $MAPbCl_3$ and 4-Mpy, which heightens the Raman signals of 4-Mpy molecule remarkably (EF = $2.6 \times 10^5$) [31]. Similarly, *Yang et al.* reported that the $Ta_2O_5$ substrate exhibits a remarkable SERS sensitivity with a low detection limit of $9 \times 10^{-9}$ M for methyl violet (MV) molecules by defect engineering [19]. In order to examine the interaction between $Ta_2O_5$ and the adsorbed MV molecule, the ultraviolet-visible absorption spectrum of MV-$Ta_2O_5$ complexes was collected to compare with those for neat MV and $Ta_2O_5$. The new broadened absorption peaks of the MV-$Ta_2O_5$ complexes were located in the range of 500-600 nm and indicated CT in this region, which was very close to the excitation wavelength (532 nm). Consequently, under irradiation with 532 nm, the substrate shows the strongest SERS activity with the quasi-resonance condition $\lambda_{CT} \approx \lambda_{laser}$ met. However, the $Ta_2O_5$ substrate presents poor SERS performance under two different excitation lasers (633 and 785 nm), which is most likely caused by the mismatch between the excitation laser and the molecular and CT frequencies. In addition, some plasmon-free metallic materials also show ultrasensitive SERS sensing capacity, which involves in the CT process between the energy level of molecules and the Fermi level of metallic materials [25].

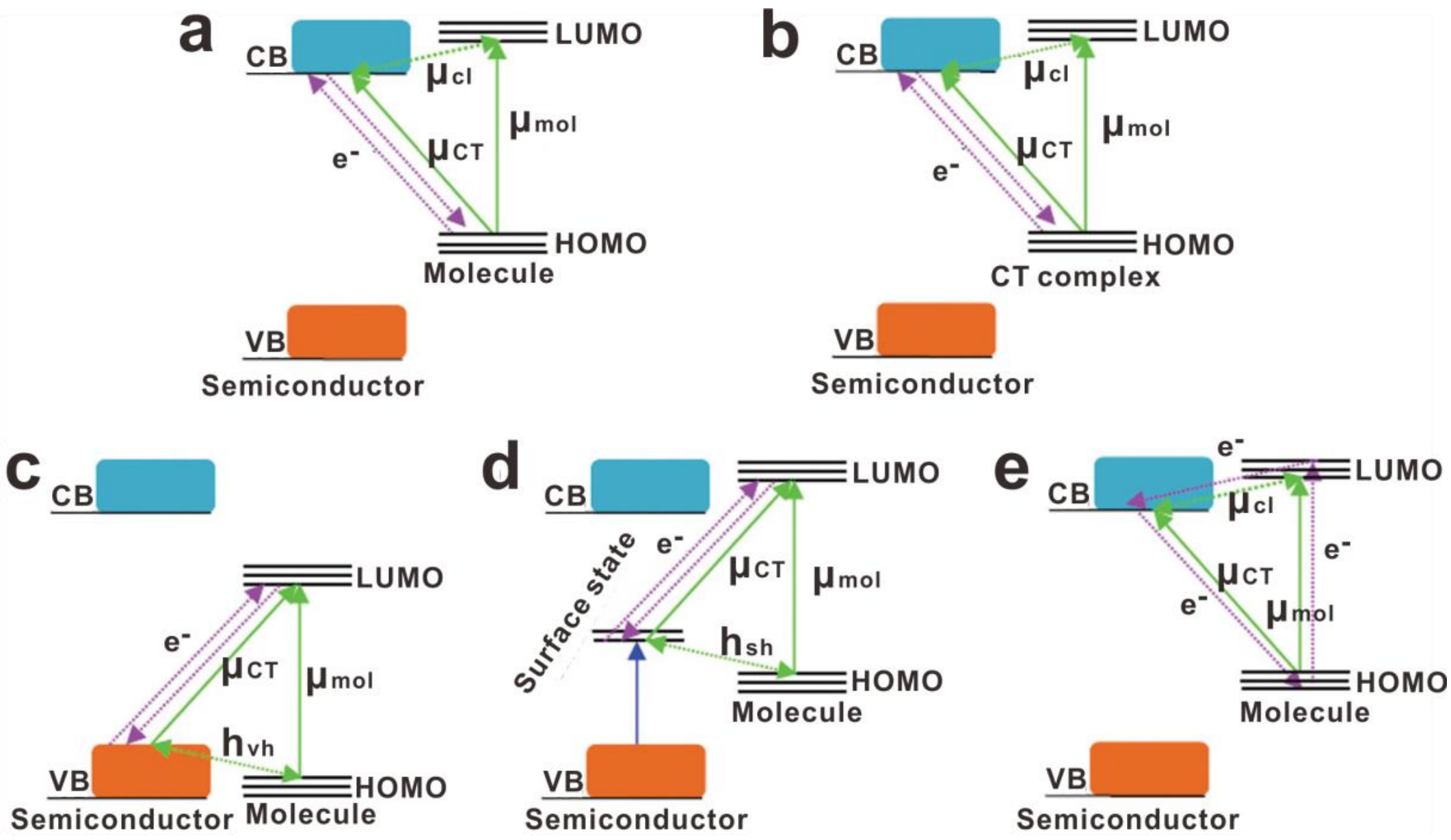

**Fig. 2** The CT pathways in semiconductor-molecule systems (a) HOMO-to-CB, (b) CT complex-to-CB, (c) VB-to-molecule LUMO, (d) surface state-to-molecule LUMO, and (e) CB-to-molecule HOMO. (Reprinted and adapted with permission from ref. [62]. Copyright 2017 The Royal Society of Chemistry.)

## 2.3 Molecular resonance and exciton resonance

The energy levels of molecules are divided into HOMO and LUMO. According to *Albrecht* and *Lombardi'* works [92-94], the $\alpha_{\sigma\rho}$ also can be expressed as $\alpha_{\sigma\rho}$ = A + B + C, where A term is relevant to the resonance Raman scattering, which follows the Franck-Condon selection rules and therefore only totally symmetric Raman modes are enhanced. The normal mode ($Q_k$) is determined by B or C term, which stems from the substrate-to-molecule and molecule-to-substrate CT transitions, respectively. Both totally and non-totally symmetric Raman modes can be enhanced by the B and C stems. And the resonance Raman scattering could be achieved when the laser excitation energy is close to the electronic transition energy of the molecule [59]. Further, according to the Herzberg-Teller coupling theory, the Raman scattering intensity contributed by the coupling of molecular resonance and charge transfer resonance is proportional to $|R_{mol\text{-}CT}(\omega)|^2$. Its expression can be described as [94]:

$$R_{mol-CT}(\omega)=\frac{(\mu_{mol}\cdot E)(\mu_{CT}\cdot E)h_{mol-CT}\langle i|Q_k|f\rangle}{((\omega_{Mie}{}^2-\omega^2)+\gamma_{Mie}{}^2)((\omega_{CT}{}^2-\omega^2)+\gamma_{CT}{}^2)((\omega_{mol}{}^2-\omega^2)+\gamma_{mol}{}^2)} \quad (5)$$

wherein Mie resonance ($\omega = \omega_{\mathrm{Mie}}$) can be achieved by adjusting the size of particles. When $\omega = \omega_{\mathrm{CT}}$ or $\omega = \omega_{\mathrm{mol}}$, CT resonance or molecular resonance can be realized, respectively. The molecular resonance and CT resonance can be couple to each other through the Herzberg-Teller coupling constant ($h_{\text{ex-mol}}$). For instance, the resonances for rhodamine 6G (R6G) and MV molecules are both located in the visible region, which can well match with the wavelength of ordinary commercial lasers. Thus, R6G and MV have been widely exploited on the plasmon-free substrates to obtain ultrasensitive SERS sensing by coupling resonance [19, 64].

Exciton resonance is often ignored in the metal SERS field. However, in view of its crucial rule in semiconductor spectroscopy, the effect of excition resonance on semiconductor SERS spectrum should be considered. In the band structure of semiconductor, the VB and CB are separated by the band gap, similar to HOMO and LUMO of molecules. The electron can be excited from the VB to the CB under optical excitation and simultaneously a hole is created in the VB, thus forming an electron-hole pair. Electron and hole possess negative and positive charge respectively, so they are attracted to each other by coulomb forces. This strongly correlated electron-hole pair is called exciton and the exciton Bohr radius is considered as the distance in an electron-hole pair. The excition resonance contributes to SERS in the same way as the molecular resonance, *i.e.* as a source of intensity borrowing. Thus, a new term involving the excition resonance in place of the molecular resonance can be described as [94]:

$$R_{ex-CT}(\omega)=\frac{(\mu_{ex}\cdot E)(\mu_{CT}\cdot E)h_{ex-CT}\langle i|Q_k|f\rangle}{((\omega_{Mie}{}^2-\omega^2)+\gamma_{Mie}{}^2)((\omega_{CT}{}^2-\omega^2)+\gamma_{CT}{}^2)((\omega_{ex}{}^2-\omega^2)+\gamma_{ex}{}^2)} \quad (6)$$

the CT resonance or excition resonance can be realized when $\omega = \omega_{\mathrm{CT}}$ or $\omega = \omega_{\mathrm{ex}}$. They are couple to each other through the Herzberg-Teller coupling constant ($h_{\text{ex-CT}}$). For example, *Lombardi et al.* measured the SERS signals of 4-Mpy molecules absorbed on CdSe quantum dots (QDs) with different sizes, and a maximum EF as high as $10^5$ was obtained in 5 nm CdSe QDs [95]. Further, they demonstrated that the exciton resonance plays a crucial role in Raman enhancement because the energy of incident light is close to exciton transition energy in 5 nm CdSe QDs. A vibrational coupling of the CT transition and exciton transition in CdSe-4-Mpy system leads to strongest SERS effect. Besides, our group fabricated a series of few-layer $MnPS_{3-x}Se_x$ ($0 \leq x \leq 3$). Among these 2D materials, the $MnPS_{2.4}Se_{0.6}$ shows the largest Raman enhancement with a LOD of $10^{-9}$ M because its exciton resonance at 2.20 eV is very close to the energy of incident light (2.32 eV). This result indicates that the synergistic resonances between CT and exciton resonance are crucial to Raman enhancement in $MnPS_{2.4}Se_{0.6}$- R6G system (Fig. 3) [96].

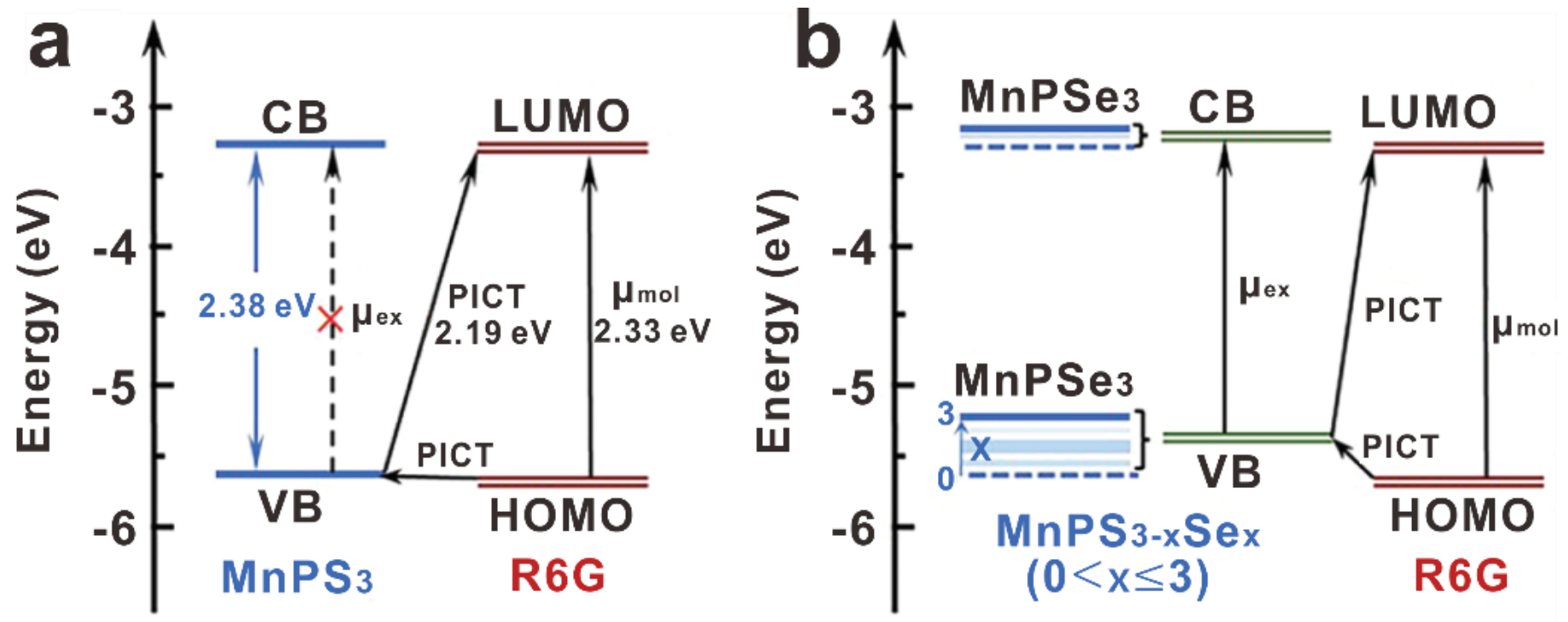

**Fig. 3** Schematic illustration of the Raman enhancement mechanism for (a) $MnPS_3$- and (b) $MnPS_{3-x}Se_x$ ($0 < x \leq 3$)-R6G system. (Reprinted and adapted with permission from ref. [96]. Copyright 2020 Wiley-VCH.)

## 3 The strategies for improving the sensitivity of plasmon-free SERS

In the past few decades, the inferior detection sensitivity impedes the practical applications of plasmon-free SERS. Fortunately, some recent studies have made remarkable progress in obtaining ultrasensitive plasmon-free SERS. A series of new strategies including defect engineering, molecular engineering, phase engineering, facet engineering, amorphization treatment, heterojunction, and morphology design have been used to improve the sensitivity of plasmon-free SERS, providing cases for understanding its enhancement mechanisms. It is distinct from the noble metals that several strategies can be applied simultaneously in plasmon-free SERS materials to further improve the SERS performance.

### 3.1 Defect engineering

As a crucial research object in materials science, defect is closely related to many properties of material. Specifically, the crystal structure alters significantly when induced defects, such as forming vacancies or dislocations in the lattice, coinciding with changes in physical and chemical properties [97-99]. Based on this point, defect engineering has been extensively applied in catalysis, energy storage, photoelectric detection and so on [100,101]. Recent years, by introducing defect engineering into SERS, researchers have realized the transformation from non-SERS activity to SERS activity successfully [17, 18, 24]. Thus, defect engineering has become an important strategy of adjusting plasmon-free SERS performance.

Served as one of the most widely studied SERS materials, the VB and CB of metal oxides are mainly contributed by the 2p orbital of oxygen and d orbital of metal atom respectively. Constructing defect in the metal oxides not only affects the position of VB, CB as well as the Fermi level, but also narrows the band gap through forming a defect energy level that provides an extra channel for the CT from the VB to the CB. The band structure changes are finally reflected in the absorption characteristics. Typically, the color of the tungsten oxide is turned from light yellow to purple or blue when oxygen vacancies inserted. It is worth noting that the band gap of metal oxides is generally higher than the excitation energy of Raman lasers, which seriously impedes the photo-induced charge transfer (PICT) between the metal oxide substrates and probe molecules, resulting in poor SRES performance. While the induced defects endow the metal oxides with narrower band gap and extra defect levels, thus broaden the CT access and facilitate PICT progress, which amplifies Raman scattering cross section significantly. Since the metal atoms in metal oxides are easily to be reduced with oxygen atoms missing in the crystal lattice, creating oxygen vacancies is the most common strategy for defect engineering. In this way, a series of high active SERS substrates including $WO_{3-x}$ [17, 64], $MoO_{3-x}$ [16, 102], $TiO_{2-x}$ [103, 104], $CrO_x$ [105], $Cu_2O$ [18] and $Ta_2O_5$ [19] were reported. For example, *Zhao et al.* prepared $WO_{3-x}$ nanowires with the detection limit of R6G down to $10^{-7}$ M, comparable to some of noble metal SERS substrates (Fig. 4a and b) [17]. However, more defects do not mean higher SERS activities. By studying $MoO_{3-x}$ with different oxygen vacancy concentrations, *Wu et al.* explored the relationship between the defect concentration and their SERS performance [102]. The Raman intensity mounts as the concentration increases initially. After reaching the maximum, the intensity dives until a stable value emerges. The mechanism behind this phenomenon is supposed that the different of CB, VB and Fermi level positions under different oxygen concentrations would change the relative position of molecular LUMO or HOMO levels towards $MoO_{3-x}$, ultimately resulting various SERS activities. Similar phenomenon has been observed on the $TiO_2$ SERS substrate doped with metal ions [106].

Apart from metal oxides, defect engineering could also regulate the SERS activity of the non-metal-oxide materials. *Zhao et al.* inserted oxygen atoms into the $MoS_2$ and obtained the narrower band gap and more electron states near the band edge. In consequence, the possibility of CT in the system would rise obviously under the illumination of laser, which further affects the CT between substrates and adsorbed molecules via vibration coupling. Compared to the samples without oxygen incorporation, the SERS activity of $MoS_2$ with proper oxygen incorporation is proven to be $10^5$ times larger (Fig. 4c and d) [24]. Besides, *Zou et al.* found the sulfur vacancies in monolayer $MoS_2$ would bring in extra electron states, which increases the EF up to 6.4 times [107]. By doping Ni into ZnS nanocrystals, the Raman intensity of 4-mercaptobenzoic acid (4-MBA) molecules detected on ZnS is magnified significantly [108]. Moreover, by controlling nitrogen doping, the Fermi level of graphene could be shifted and align with the LUMO of the molecule, significantly amplifying the vibrational Raman modes of the molecule. Thus, the nitrogen doped graphene showed an ultralow LOD of $10^{-11}$ M [34].

The technical approaches of creating defects in the SERS materials are roughly divided into two categories. On the one hand, we can adjust the synthesis procedure to produce defects directly. As a typical example, *Guo et al.* fabricated defective $Cu_2O$ SERS substrates by the means of chemical precipitation method, on which the LOD is as low as $10^{-9}$ M and the corresponding EF reaches $10^5$ [18]. In our previous work, we fabricated $MoO_{3-x}$ nanosheet ink by a solvothermal method and employed it to print the paper-based semiconducting SERS substrates successfully [16]. *Yang et al.* prepared Zn-doped $TiO_2$ nanoparticle SERS substrates by using a sol-hydrothermal method and explored the influence of Zn ions concentration on the SERS activity [109]. On the other hand, the secondary treatment for prefabricated materials could also accomplish defect engineering. The approaches general used include chemical reducing agent treatment, plasma processing, electrochemical reduction and so on [64, 110-112]. By reducing the $WO_3$ film via hydrogen gas, *Fan et al.* obtained the detection limit of R6G as low as $10^{-9}$ M [64]. And the limit of $5 \times 10^{-8}$ M was realized on the metal oxide films irradiated with argon/nitrogen ion by *Zheng et al*. [110]. Through electrochemical reduction method, *Cong* and co-workers inserted metal ions into sputtered metal oxide films and prepared highly uniform SERS substrates with excellent performance [111].

In a word, as an effective strategy to improve SERS performance of plasmon-free materials, defect engineering breaks the limit of traditional noble metals and broadens the applications of plasmon-free compounds as SERS substrates. Table 1 shows the summary of the reported plasmon-free SERS substrates obtained by defect engineering in recent years.

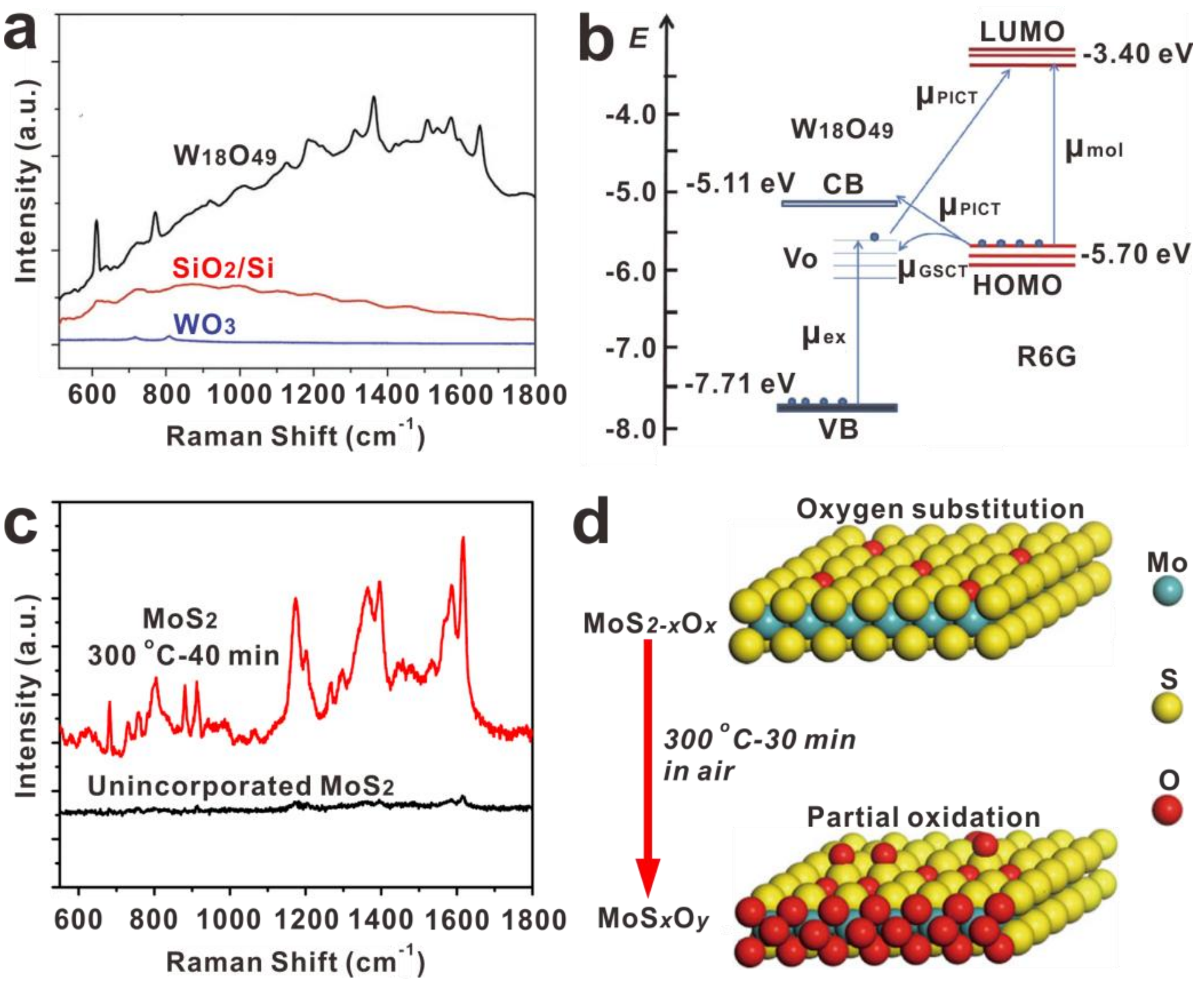

**Fig. 4** Ultrasensitive SERS substrates are obtained by using defect engineering on $W_{18}O_{49}$ and $MoS_xO_y$. (a) SERS spectra of $10^{-6}$ M R6G adsorbed on $W_{18}O_{49}$, $WO_3$ and bare $SiO_2$/Si substrates. (b) Energy-level diagram of R6G on oxygen-deficit $W_{18}O_{49}$ measured in a vacuum. (c) SERS spectra of Victoria blue B on the partially oxidized $MoS_2$ sample at 300 ℃ for 40 min and unincorporated $MoS_2$. (d) Structure of oxygen-substituted and oxidized $MoS_2$. (Reprinted and adapted with permission from (a and b) ref. [17] and (c and d) ref. [24]. Copyright 2015 and 2017 Springer Nature.)

**Table 1** The summary of the reported plasmon-free SERS substrates obtained by defect engineering in recent years.

| Substrate | Probe | LOD (M) | EF | Ref. |
|---|---|---|---|---|
| $W_{18}O_{49}$ nanowire | R6G | $10^{-7}$ | $3.4 \times 10^5$ | [17] |
| $WO_{3-x}$ film | R6G | $10^{-9}$ | $1.16 \times 10^6$ | [64] |
| $WO_{3-x}$ film | R6G | $5 \times 10^{-8}$ | $1.1 \times 10^4$ | [110] |
| $Li_xWO_3$ film | R6G | $10^{-6}$ | $8.86 \times 10^4$ | [111] |
| $W_{18}O_{49}$ nanowire film | Rhodamine B (RhB) | $10^{-7}$ | $4.38 \times 10^5$ | [113] |
| Mo doped $Ta_2O_5$ nanowire | MV | $9 \times 10^{-9}$ | $2.2 \times 10^7$ | [19] |
| Ta doped $TiO_2$ nanowire | Methylene blue (MB) | $< 2 \times 10^{-5}$ | N/A | [114] |
| Ni doped $TiO_2$ photonic microarray | 4-MBA | $1 \times 10^{-11}$ | $3.3 \times 10^{4}$ | [115] |
| Co doped $TiO_2$ nanoparticle | 4-MBA | $< 10^{-3}$ | N/A | [116] |
| Co doped $TiO_2$ nanoparticle | 4-MBA | $< 10^{-3}$ | N/A | [117] |
| Mn doped $TiO_2$ nanoparticle | 4-MBA | $< 10^{-3}$ | N/A | [106] |
| Zn doped $TiO_2$ nanoparticle | 4-MBA | $< 10^{-3}$ | N/A | [109] |
| C doped $TiO_2$ microparticle | 4-MBA | $< 10^{-3}$ | $6.2 \times 10^{3}$ | [118] |
| N doped $TiO_2$ nanofibre | 4-MBA | $< 10^{-3}$ | N/A | [119] |
| Black $TiO_2$ nanowire | R6G | $1 \times 10^{-7}$ | $1.2 \times 10^{6}$ | [103] |
| $TiO_{2-x}$ QDs | Crystal violet (CV) | $10^{-9}$ | $\sim 10^{10}$ | [104] |
| $TiO_2$ nanoparticle | 4-MBA | $1 \times 10^{-8}$ | N/A | [120] |
| $CrO_x$ film | R6G | $10^{-9}$ | N/A | [105] |
| $MoO_{3-x}$ film | R6G | $10^{-7}$ | N/A | [105] |
| $MoO_{3-x}$ micron urchin | R6G | $10^{-7}$ | $\sim 10^{5}$ | [121] |
| $MoO_{3-x}$ nanobelt | R6G | $10^{-8}$ | $1.8 \times 10^{7}$ | [102] |
| $MoO_{3-x}$ nanosheet | R6G | $10^{-7}$ | $3.32 \times 10^{5}$ | [16] |
| $V_2O_{5-x}$ nanoparticle | R6G | $10^{-6}$ | N/A | [102] |
| $V_2O_5$ nanoparticle | R6G | $10^{-8}$ | N/A | [122] |
| Nd doped ZnO nanoparticle | Malachite green | $10^{-7}$ | N/A | [123] |
| Nd doped ZnO micron urchin | 4-Mpy | N/A | N/A | [124] |
| Ga doped ZnO nanoparticle | 4-MBA | $< 1 \times 10^{-3}$ | $8.13 \times 10^{3}$ | [125] |
| Ni doped ZnO nanoparticle | 4-MBA | $< 1 \times 10^{-3}$ | N/A | [126] |
| Co doped ZnO nanoparticle | 4-MBA | $< 1 \times 10^{-3}$ | N/A | [127] |
| Mg doped ZnO nanocrystal | 4-MBA | N/A | N/A | [128] |
| ZnO QDs | CV | $10^{-9}$ | $\sim 10^{6}$ | [129] |
| ZnO nanosheet | 4-Mpy | $1 \times 10^{-7}$ | $7.7 \times 10^{5}$ | [130] |
| ZnO nanoparticle | Methyl orange | $< 1 \times 10^{-3}$ | 56 | [131] |
| $Cu_2O$ Superstructure particle | R6G | $10^{-9}$ | $8 \times 10^{5}$ | [18] |
| $SnO_2$ nanoparticle | 4-MBA | $<1\times10^{-3}$ | $3.4\times10^{3}$ | [132] |
| O doped 2H-$MoS_2$ nanosheet | R6G | $10^{-7}$ | $1.4\times10^{5}$ | [24] |
| O doped 1T-$MoS_2$ nanosheet | R6G | $10^{-11}$ | $1.24\times10^{7}$ | [133] |
| O doped 2D-$MoS_2$ | R6G | $<10^{-5}$ | N/A | [134] |
| $MoS_{2-x}$ monolayer | R6G | $<10^{-6}$ | N/A | [107] |
| $Cu_{2-x}S$ nanoparticle | R6G | $10^{-7}$ | $3.4\times10^{4}$ | [135] |
| Zn doped $ZrO_2$ nanoparticle | 4-MBA | $<1\times10^{-3}$ | $1.94\times10^{4}$ | [136] |
| Indium tin oxide (ITO) film | R6G | $<10^{-5}$ | N/A | [137] |
| ITO film | R6G | $1\times10^{-7}$ | N/A | [138] |
| Ni doped ZnS nanoparticle | 4-MBA | $<1\times10^{-3}$ | N/A | [108] |
| 2D Sb:Na | R6G | $10^{-16}$ | N/A | [139] |
| ZnSe nanowire | R6G | $10^{-11}$ | $6.12\times10^{7}$ | [140] |
| N doped Graphene | R6G | $10^{-11}$ | N/A | [34] |
| N doped Graphene QDs | RhB | $10^{-10}$ | $3.2\times10^{3}$ | [37] |
| Fe-N-doped carbon nanosheet/rod | R6G | $1\times10^{-7}$ | $3.8\times10^{4}$ | [141] |
| B doped diamond film | MB | $1\times10^{-7}$ | $3.2\times10^{5}$ | [142] |
| Reduced $Ti_3C_2T_x$ nanosheet | R6G | $1\times10^{-7}$ | $1.0\times10^{7}$ | [52] |
| Reduced $MnCo_2O_4$ nanotube | CV | $1\times10^{-5}$ | $1.34\times10^{6}$ | [143] |
| S doped $SnSe_2$ nanosheet | R6G | $1\times10^{-7}$ | N/A | [144] |
| $WSe_{2-x}$ monolayer | CuPc | N/A | 120 | [112] |

### 3.2 Amorphization treatment

Although amorphous materials lack long-range atomic order, there still retain the basic building blocks and short-range order over a few atoms. For the crystal materials, electrons do not have absolutely free character due to their highly ordered periodic lattices. Nevertheless, the long-range unordered structure of the amorphous materials will produce the surface suspension bonds and band tails, which can promote the escape and transfer of surface electrons, and cause novel physical and chemical phenomena [145]. Recently, it has attracted increasing attention to introduce the amorphous materials into SERS researches [64-66]. For instance, *Wang et al.*

designed a novel amorphous ZnO (a-ZnO) nanocage and the EF of mercapto molecules on these a-ZnO nanocages was up to 6.62 $\times 10^5$ (Fig. 5a and b) [65]. First-principle density functional theory (DFT) calculation clearly confirmed that a-ZnO nanocages enabled more facile and efficient CT from the ZnO surface to the probe molecules compared to their crystalline counterparts. When mercapto molecules were adsorbed on the a-ZnO nanocages surface, the highly effective CT effect could even form the π bonding in the Zn-S bonds, which was further verified by the X-ray absorption near-edge structure (XANES) characterization. The metastable electronic states of a-ZnO nanocages surface impose a weaker constraint to the surface electrons, which could contribute to the efficient CT process in a-ZnO-molecule system, leading to significant magnification of the molecular polarization and remarkable enhancement of Raman scattering. Furthermore, they fabricated novel 2D amorphous $TiO_2$ (a-$TiO_2$) nanosheets, which exhibited a remarkable SERS sensitivity with an ultrahigh EF of 1.86 $\times 10^6$, stronger than that of their 2D crystalline $TiO_2$ (c-$TiO_2$) nanosheets counterparts (Fig. 5c-e) [66]. According to the results of DFT calculations, the band gap of 4-MBA@a-$TiO_2$ complex was only 0.729 eV, smaller than that of the 4-MBA@c-$TiO_2$ complex (2.661 eV). As the HOMO-LUMO energy gap of 4-MBA was about 4.112 eV, the obviously reduced band gap of the 4-MBA@a-$TiO_2$ complex could be ascribed to the strong energy level coupling between the 4-MBA molecule and the 2D a-$TiO_2$, which allowed more possible thermodynamic PICT excitations at the low-energy level and hence effectively enhanced the Raman signals. Besides, DFT calculations and experimental results further revealed that the 2D a-$TiO_2$ nanosheets possessed smaller band gap and higher electronic density of states (DOS) than that of their crystal counterparts, which further effectively enhanced the vibronic coupling of resonances in the a-$TiO_2$-molecule system, resulting in a high-efficient PICT process. In addition, our group also observed ultrasensitive SERS activity in amorphous tungsten oxide (a-$WO_{3-x}$) films with a high EF of 1.16 $\times 10^6$ [64]. Compared with crystalline $WO_{3-x}$ (c-$WO_{3-x}$) films, a-$WO_{3-x}$ films exhibited narrower band gap, stronger exciton resonance, and higher DOS near the Fermi level, which could promote the PICT resonance between analytes and substrates by offering efficient routes of charge escaping and transferring as well as strong vibronic coupling, thus realizing ultrasensitive SERS activity on a-$WO_{3-x}$ films (Fig. 5f). Besides, researchers have also observed strong SERS signals from some amorphous materials, such as $Rh_3S_6$ [146], $M(OH)_x$ (M = Fe, Co, Ni and Zn) [147, 148], $Nb_2O_5$ [149]. However, for the future development of amorphous material-based SERS substrates, there are still some obstacles to be settled. For example, the present preparation process of amorphous SERS materials is pretty complicated thus a new convenient and controllable synthetic method is urgently needed. And it is still a challenge to optimize the energy band structure and surface physicochemical properties of amorphous SERS materials for the programmable control of CT pathways and coupling enhancement.

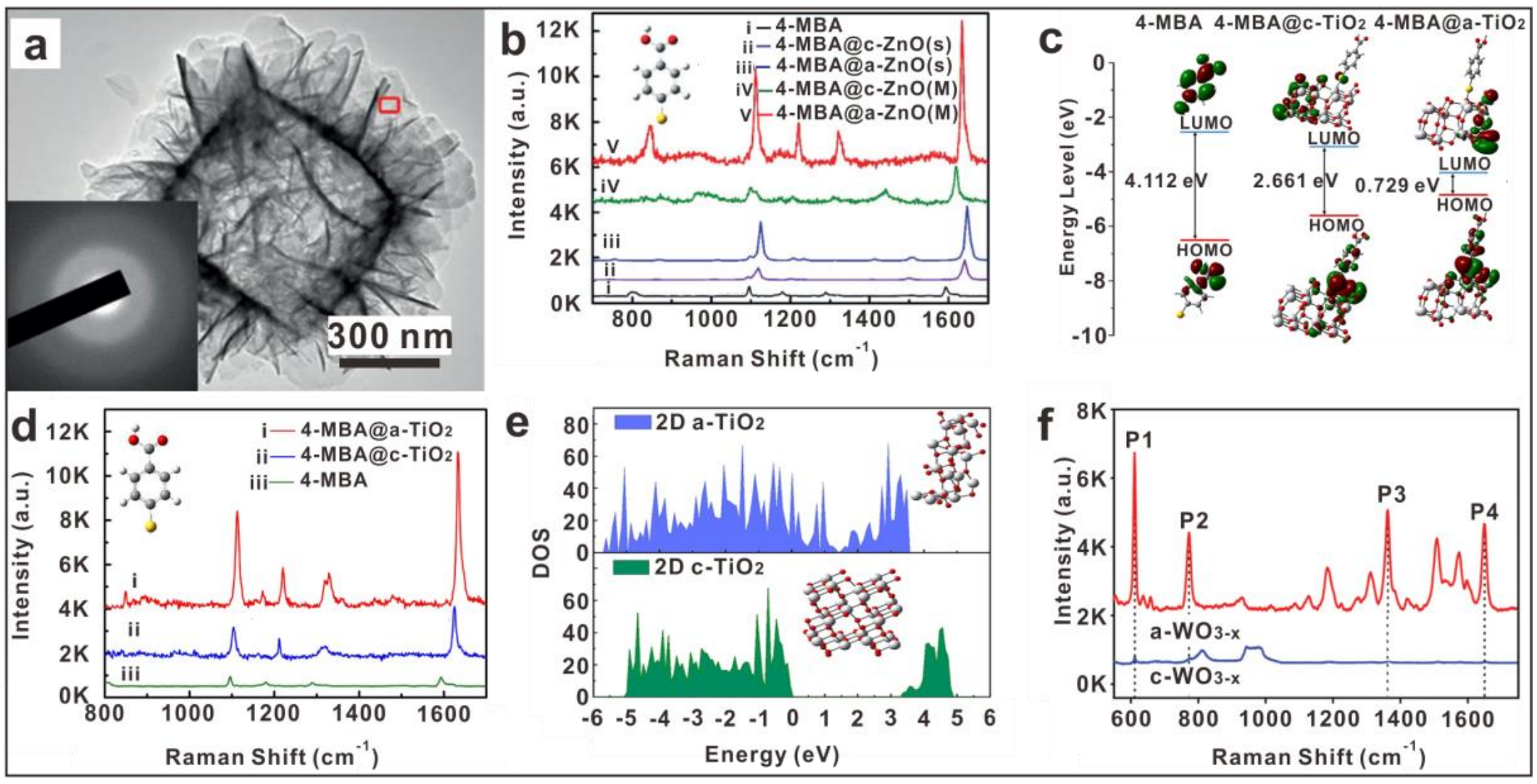

**Fig. 5** Ultrasensitive SERS substrates are obtained by using amorphization treatment on ZnO, $TiO_2$ and $WO_{3-x}$. (a) SEM image of a-ZnO nanocages. (b) Measured (M) and simulated (S) SERS spectra of 4-MBA adsorbed onto a single a-ZnO nanocage. (c) Schematic energy level diagrams in the 4-MBA-$TiO_2$ system. (d) SERS spectra of 4-MBA adsorbed on a- and c-$TiO_2$ nanosheets. (e) DOS calculation of 2D a- and c-$TiO_2$. (f) SERS spectra of R6G adsorbed on a- and c-$WO_{3-x}$ films. (Reprinted and adapted with permission from (a and b) ref. [65], (c-e) ref. [66], and (f) ref. [64]. Copyright 2017 Wiley-VCH. Copyright 2019 American Chemical Society. Copyright 2019 Wiley-VCH.)

### 3.3 Phase engineering

TMDs possess complex and variable phase structures and electronic structures. Taking $MoS_2$ as an example, the steady phase is hexagonal structure (2H phase), belonging to semiconductor. While once the position of coordinating atoms in 2H-$MoS_2$ is induced to be altered, metastable octahedral structure is formed, which is called metallic 1T-$MoS_2$ [150]. Recently, metallic/semimetallic TMDs have attracted much attention due to their unique physical and chemical properties [151, 152]. More importantly, metallic/semimetallic TMDs possess large DOS near the Fermi levels and high surface activities, indicating that they can provide strong molecule-SERS substrate coupling as well as effective CT, therefore they may become promising candidates for high-performance SERS materials. *Yin et al.* reported the SERS properties of liquid exfoliation-obtained monolayer 2H-$MoS_2$, 2H-$MoSe_2$, 1T-$MoS_2$ and 1T-$MoSe_2$ (Fig. 6a-e) [27]. When the phase was transformed form 2H- to 1T-phase, the Raman enhancement of molecules adsorbed on $MoX_2$ (X=S, Se) monolayers can increase significantly. For semiconducting 2H-$MoX_2$, electrons can transfer from the VB of 2H-$MoX_2$ to HOMO level of CuPc molecules with enough extra energy. Compared with 2H-$MoX_2$, metallic 1T-$MoX_2$ possesses larger available DOS between the levels of HOMO and Fermi energy, which induces highly efficient CT from the Fermi energy level of 1T-$MoX_2$ to HOMO level of CuPc without requiring extra energy. Thus, larger enhancement in SERS was found in metallic 1T-$MoX_2$. Besides, *Tao et al.* confirmed that atomic layers of semimetallic 1T′-W(Mo)$Te_2$ can significantly enhance the signals of the R6G and achieve a remarkable LOD of $10^{-14}$ ($10^{-13}$) M (Fig. 6f-g) [25]. For 1T′-W(Mo)$Te_2$, the EM contribution is excluded since the LSPR of W(Mo)$Te_2$ is located in the mid-infrared region. The strong interaction between the analyte and 1T′-W(Mo)$Te_2$ as well as abundant DOS near the Fermi level of 1T′-W(Mo)$Te_2$ lead to the prominent SERS effect by promoting the CT resonance in the analyte-W(Mo)$Te_2$ system. Other metallic/semimetallic TMDs SERS materials such as $ReS_2$ [28] and $NbS_2$ [26] have also been reported. In addition, *Miao et al.* demonstrated that the phase transition induced Raman enhancement on $VO_2$ nanosheets [153]. During the high temperature annealing process, monoclinic $VO_2$(B) could transform to another monoclinic $VO_2$(M) and to highly ordered tetragonal rutile $VO_2$(R). Among these phase structures, low crystal symmetry of $VO_2$(B) exhibits highest SERS activity and achieves a $10^{-7}$ M level detection limit for various dye molecules.

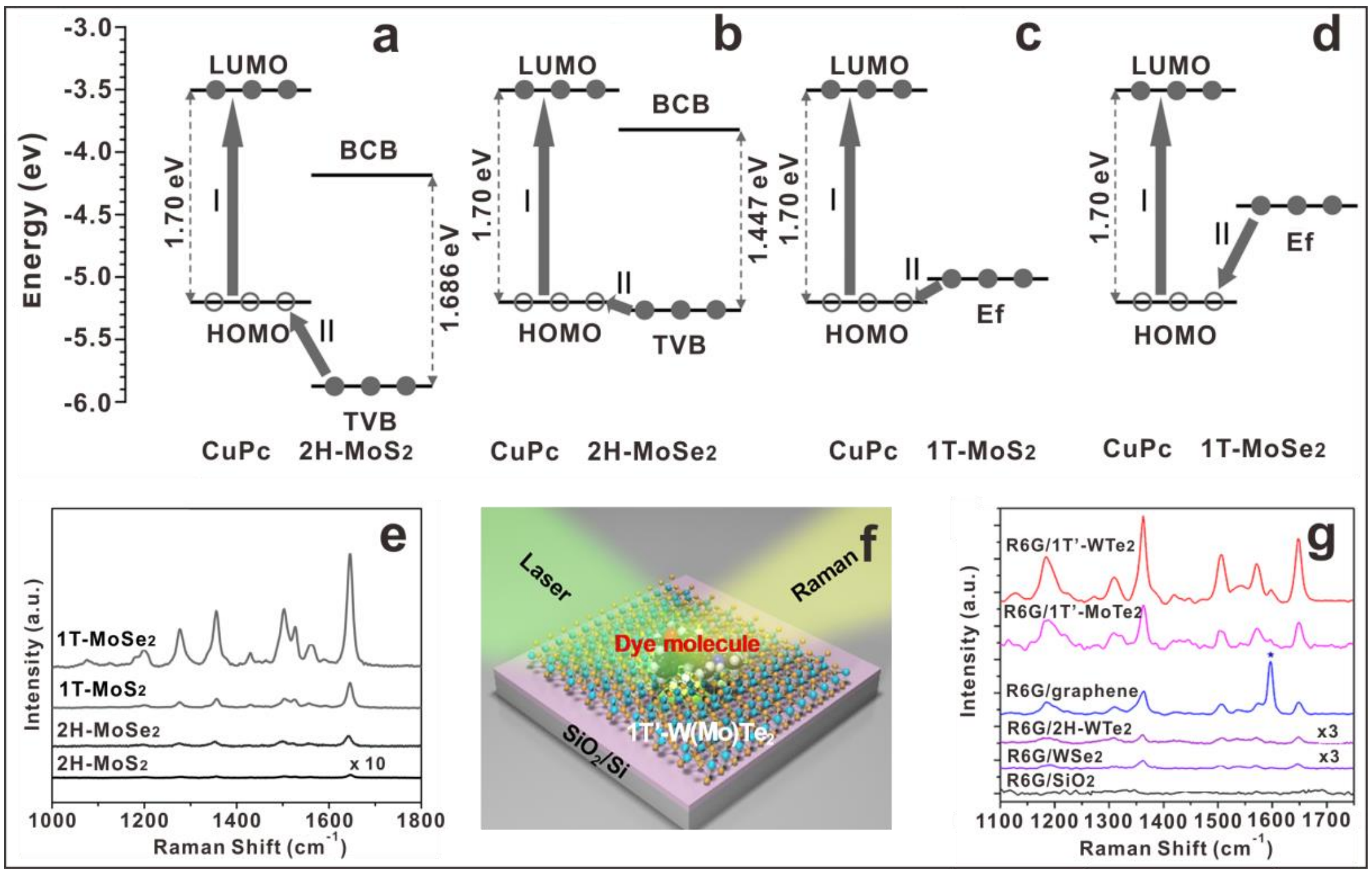

**Fig. 6** Ultrasensitive SERS substrates are obtained by using phase engineering on $MoX_2$ (X=S, Se) and W(Mo)$Te_2$. (a-d) Schematic illustration of the energy band diagrams and CT process from 2H/1T-$MoX_2$ (X = S, Se) monolayer to CuPc. (e) SERS spectra of R6G adsorbed on 2H/1T-$MoX_2$ (X = S, Se) monolayer. (f) SERS effects on the 1T′-W(Mo)$Te_2$. (g) SERS spectra of R6G coated on 1T′-W(Mo)$Te_2$ and other substrates. (Reprinted and adapted with permission from (a-e) ref. [27], and (f and g) ref. [25]. Copyright 2017 Wiley-VCH. Copyright 2018 American Chemical Society.)

### 3.4 Heterojunction engineering

Heterojunction engineering has drawn intense research attention because it can provide multifunctional and enhanced properties that cannot be achieved by the individual material [154-157]. Recently, researchers introduce heterojunction engineering into the plasmon-free SERS fields, providing a brand-new way to obtain ultrasensitive plasmon-free SERS substrates [61, 67, 68, 158]. For example, our group designed a $W_{18}O_{49}$/monolayer $MoS_2$ (W/M) vertical heterojunction as an ultrasensitive SERS substrate (Fig. 7a-c) [67]. The substrate was successfully fabricated by coating a $W_{18}O_{49}$ thin layer on chemical vapor deposition (CVD)-grown monolayer $MoS_2$ via magnetron sputtering then annealing under hydrogen atmosphere. It was confirmed that the coupling of these two materials could lead to dramatic enhancement of PICT processes. The LOD for R6G molecule on W/M substrate is as low as $1 \times 10^{-9}$ M with the maximum EF up to $3.45 \times 10^{7}$. The EM enhancement is not considered because the LSPR of $W_{18}O_{49}$ is located at the mid-infrared region. According to the Raman spectra, there is considerable Raman enhancement under the wavelengths of 532 nm, while no detectable Raman signals exist at 633 and 785 nm excitation, indicating that Raman enhancement on the W/M substrate is a selective process by the CT mechanism. On the one hand, the internal properties of the W/M substrate promote efficient separation of electron-hole pairs and charge transformations, which could significantly improve the electron density on the surface of the substrate in contact with the adsorbed molecules. On the other hand, there are more possible exciton resonance transitions in the heterojunction structure. These synergistic effects could dramatically enhance the CT processes, which could outstandingly increase the SERS sensing capacity of W/M vertical heterojunction substrates. The black $TiO_2$ nanoparticle with crystal-amorphous core-shell heterojunction has been reported as a SERS substrate by *Lin et al.* [159]. Benefitting from the synergistic effects of the novel crystal-amorphous core-shell structure, this substrate shows a LOD of $10^{-6}$ M and a high EF value of $4.3 \times 10^{5}$. High-efficiency exciton transitions in the crystal core can generate plentiful excitons, which can provide sufficient charge source. Subsequently, the heterojunction structure enables the efficient exciton separation at the interface of crystal-amorphous, which can effectively facilitate the CT from crystal core to amorphous shell and result in exciton enrichment at the amorphous shell. Besides, it was further confirmed that the amorphous shell structure possesses lower Fermi level position, narrower band gap, and higher DOS than crystal core. These features can offer effective routes of CT as well as strong vibronic coupling in $TiO_2$-analyte system, which are prominent in subsequent SERS enhancement. Similarly, other heterojunction structures, such as $MoS_2$/ZnO [160], graphene/$ReO_xS_y$ (Fig. 7d and e) [61], graphene/$WSe_2$ [68], and InAs/GaAs [161] are also investigated. These multitudinous heterojunction substrates provide excellent platforms for ultrasensitive SERS sensing. Table 2 shows the summary of the reported plasmon-free SERS substrates obtained by heterojunction engineering in recent years.

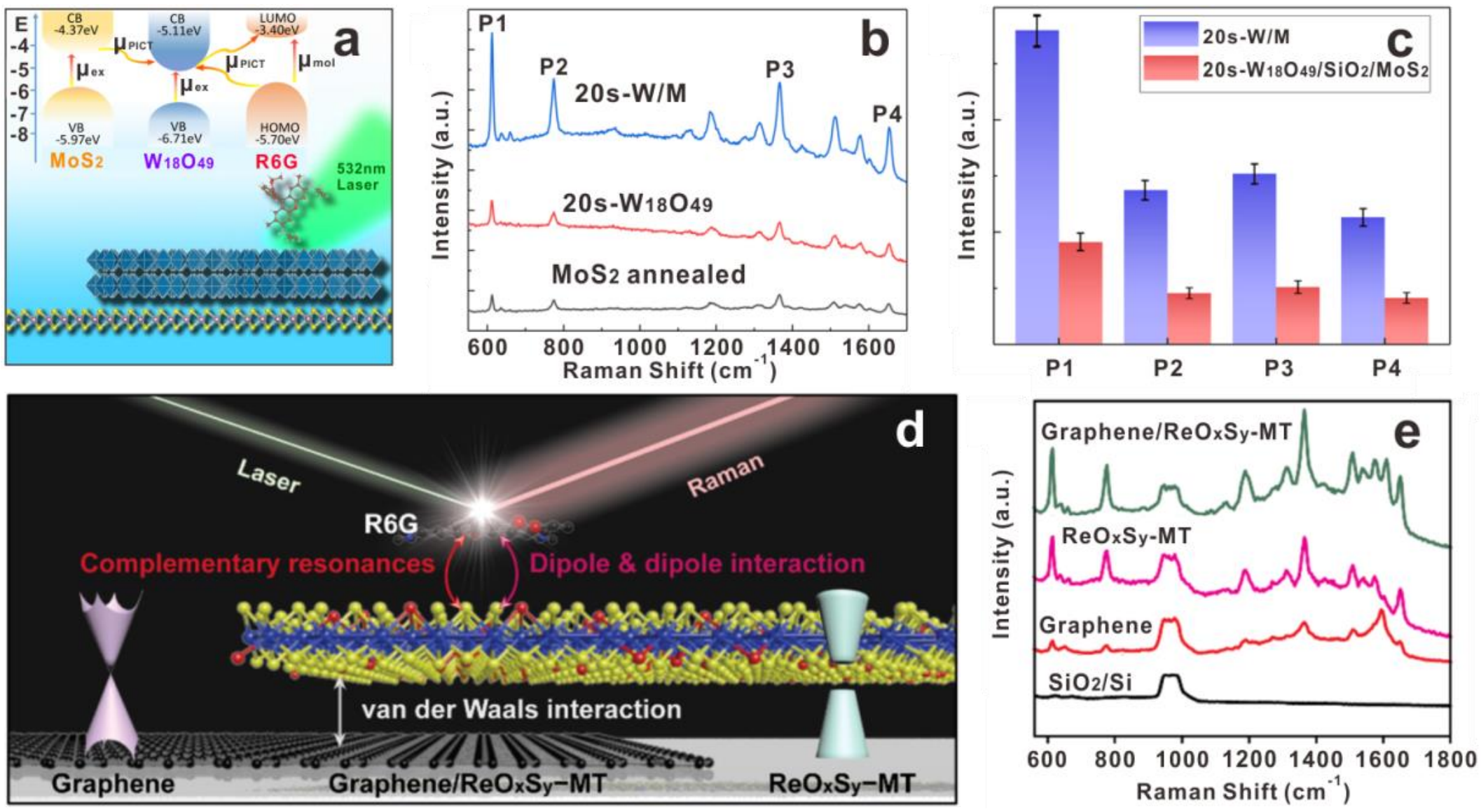

**Fig. 7** Ultrasensitive SERS substrates are obtained by using heterojunction engineering on W/M and graphene/$ReO_xS_y$. (a) Schematic energy level diagrams and CT in the R6G-W/M ternary system. (b) SERS spectra of R6G adsorbed on the annealed monolayer $MoS_2$, $W_{18}O_{49}$, and W/M substrates. (c) SERS spectral comparison chart of R6G adsorbed on the W/M and $W_{18}O_{49}/SiO_2/MoS_2$ substrates. (d) Illustration of measured Raman enhancement of R6G on graphene/$ReO_xS_y$ films. (e) SERS spectra of R6G adsorbed on graphene, $ReO_xS_y$, and graphene/$ReO_xS_y$ substrates. (Reprinted and adapted with permission from (a-c) ref. [67] and (d-e) ref. [61]. Copyright 2019-2020 American Chemical Society.)

**Table 2** The summary of the reported plasmon-free SERS substrates obtained by heterojunction engineering in recent years.

| Substrate | Probe | LOD (M) | EF | Ref. |
|---|---|---|---|---|
| 2D $W_{18}O_{49}/MoS_2$ | R6G | $1\times10^{-9}$ | $3.45\times10^{7}$ | [67] |
| $TiO_2$ core-shell nanoparticle | R6G | $10^{-6}$ | $4.3\times10^{5}$ | [159] |
| 2D graphene/$ReO_xS_y$ | R6G | $1\times10^{-14}$ | N/A | [61] |
| 2D graphene/$WSe_2$ | CuPc | N/A | 78.2 | [68] |
| 2D TMD/$SnSe_2$ | TMD | N/A | 10 | [162] |
| 1T/2H-$MoS_2$ nanosheet | R6G | $5\times10^{-8}$ | N/A | [163] |
| 1T/2H-$WS_2$ nanosheet | R6G | $5\times10^{-8}$ | N/A | [163] |
| InAs/GaAs QDs | Pyridne | N/A | $\sim10^{3}$ | [161] |
| $Ag_2S$/graphene oxide (GO) nanoparticle/sheet | $Ag_2S$/GO | N/A | N/A | [164] |
| ZnS/GO nanoparticle/sheet | 4-Mpy | $1\times10^{-5}$ | N/A | [165] |
| $SiO_2/TiO_2$ core-shell | Glutathione | $<1\times10^{-3}$ | N/A | [166] |
| GO/$TiO_2$ inverse opal | MB | $6\times10^{-7}$ | $5\times10^{4}$ | [167] |
| GO/$TiO_2$ nanocomposite | CuPc | N/A | 48.2 | [168] |
| GO/ZnO nanocomposite | R6G | $1\times10^{-10}$ | $7\times10^{4}$ | [169] |
| $TiO_2$- carbonaceous nanotube | MB | N/A | $3\times10^{3}$ | [170] |
| Reduced graphene oxide (rGO)-$TiO_2$ nanoparticle/sheet | 4-MBA | $1\times10^{-7}$ | $5.5\times10^{6}$ | [171] |
| rGO-$TiO_2$-$Fe_3O_4$ nanoparticle/sheet | 4-MBA | $1\times10^{-10}$ | $2.7\times10^{7}$ | [172] |
| $Fe_3O_4$@GO@$TiO_2$ nanoparticle | CuPc | N/A | $8.08\times10^{6}$ | [173] |
| ZnO/ZnS/$MoS_2$ nanoparticle | R6G | $1\times10^{-9}$ | $1.4\times10^{8}$ | [174] |
| $MoS_2$/ZnO nanoflower/particle | MB | $1\times10^{-12}$ | $1.13\times10^{6}$ | [175] |
| $MoS_2$/ZnO microflower/particle | Bisphenol A | $1\times10^{-9}$ | $5.8\times10^{5}$ | [176] |
| Graphene/$MoS_2$ nanoflower/sheet | Rhodamine B | $5\times10^{-11}$ | $2.96\times10^{7}$ | [177] |
| $CsPbBr_3$@ZIF-8 composite | 4-Mpy | N/A | $1.17\times10^{5}$ | [178] |

### 3.5 Facet engineering

As an effective strategy for tuning the properties of nanomaterials, facet engineering has been applied in catalysis, gas sensors, and energy storage [179-181]. Nanomaterials with the selective exposure of facets exhibit distinctive surface electronic structures and surface electron transport properties due to the diverse atomic coordination and configurations. Recently, facet engineering has been used for improving the SERS behavior of plasmon-free materials [69, 70]. For example, compared with much less affinity for the {101} $TiO_2$, *Urdaneta et al.* reported a selective adsorption of dopamine (DA) on the {001} and {100} terminations, which was consistent with the CT associated to the SERS effect [182]. *Guo et al.* synthesized three different facets of $Cu_2O$ microcrystals: {100}-cubic $Cu_2O$, {110}-octahedral $Cu_2O$, and {111}-dodecahedral $Cu_2O$ (Fig. 8a and b) [69]. Among these $Cu_2O$ microcrystals, {100} $Cu_2O$ exhibited the strongest Raman signals. In {100} $Cu_2O$, Kelvin probe force microscopy (KPFM) technology and first-principle DFT calculation evidenced the highest interfacial CT based on the facet-dependent work function. More efficient CT process in {100} $Cu_2O$ was obtained as its lowest electronic work function, resulting in the largest magnification of Raman scattering cross section. It is also worth mentioning that *Yu et al.* obtained a high EF of $1.6 \times 10^6$ from DA molecules adsorbed on a symmetric prickly {201} $TiO_2$, which is three orders of magnitude higher than that on asymmetric {001}, {101}, and {100} $TiO_2$ (Fig. 8c and d) [70]. Through combining XANES analysis, they demonstrated that the high density of unoccupied $t_{2g}$ orbitals in {201} $TiO_2$ played a crucial role in efficient CT process, which made the CT of {201} $TiO_2$ 1.3-1.8 times higher than others. In addition, electronic DOS calculations also demonstrated that the urchin-like spheres of {201} $TiO_2$ could generate a strong electromagnetic field enhancement, which further contributes to the Raman enhancement in DA-{201} $TiO_2$ system. *Zhang et al.* also reported that the SERS activity of $TiO_2$ could be improved by optimizing the exposed ratio between the {001} and {101} facets [183]. They found that the SERS performance was correlated with the enhanced mobility of photo-induced electrons on exposed {001} facets, which was promoted by the surface electric field pointing from exposed {101} facets to the junction edge, and then to {001} facets because of the heterogeneous facets potentials. So far, facet-dependent SERS materials almost are confined to metal oxides, it is possible to improve the SERS activity of non-metal-oxide materials by using facet engineering.

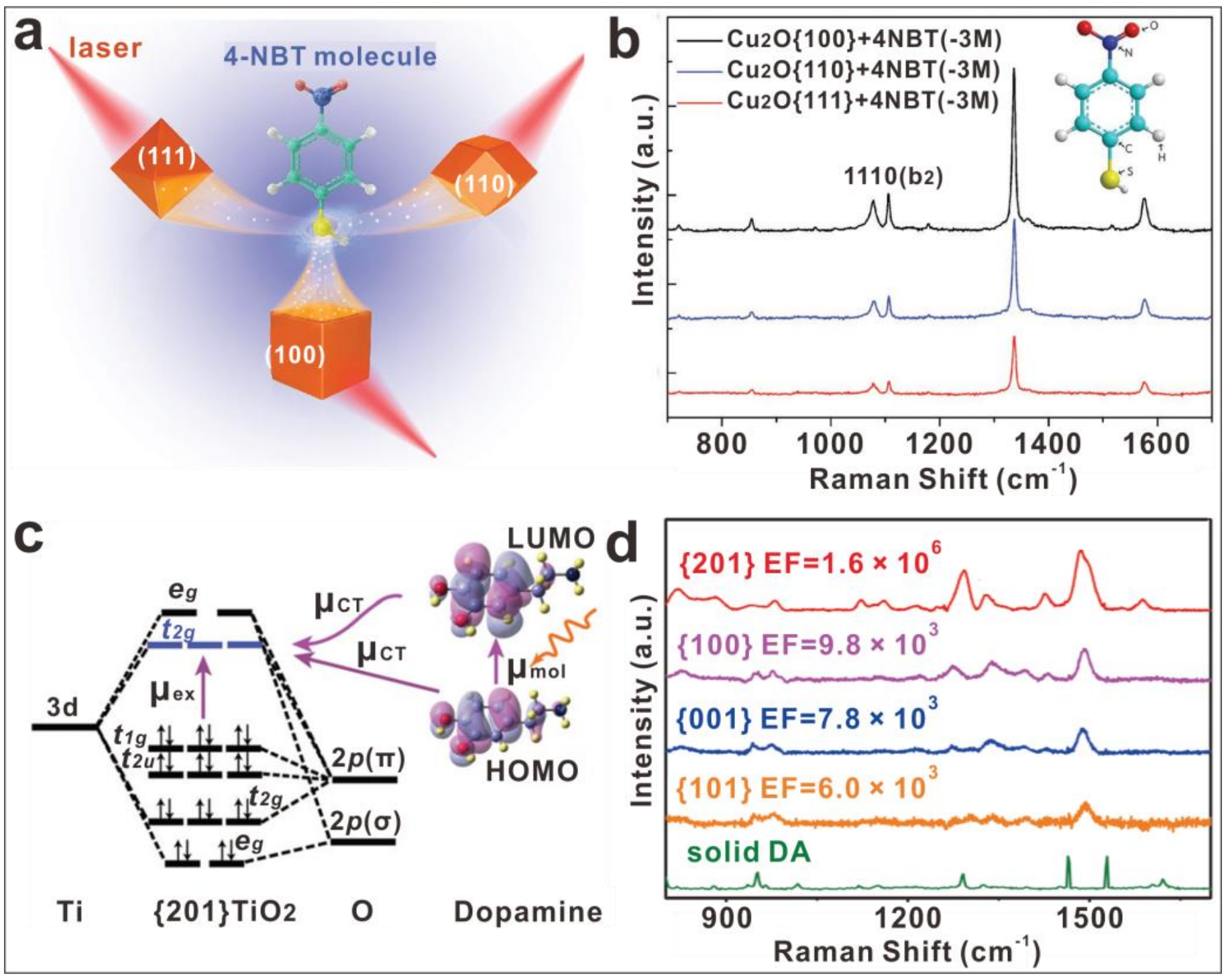

**Fig. 8** Ultrasensitive SERS substrates are obtained by using facet engineering on $Cu_2O$ and $TiO_2$. (a) Schematic illustration of CT process in the SERS effect of the three $Cu_2O$ polyhedra. (b) SERS spectra of 4-nitrobenzenethiol obtained on three $Cu_2O$ polyhedra. (c) Molecular-orbital diagram for {201} $TiO_2$ and DA molecule. (c) SERS spectra of DA on different $TiO_2$ and solid DA. (Reprinted and adapted with permission from (a and b) ref. [69], and (c and d) ref. [70]. Copyright 2018 Wiley-VCH. Copyright 2019 The Royal Society of Chemistry.)

### 3.6 Molecular engineering

Π-conjugated organic small molecules that consist of π-conjugated moieties and functional groups have been widely used in optoelectronic devices as their structural diversity, facile synthesis, and highly delocalized molecular orbitals [184]. Different from the molecules only with σ-electrons, the presence of delocalized π-electrons in π-conjugated organic semiconductors allows more exceptional charge transport or light manipulation properties. In 2017, *Yilmaz et al.* designed a novel SERS platform based on a,w-diperfluorohexylquaterthiophene (DFH-4T) thin film via an oblique-angle vapor deposition [46]. This superhydrophobic and ivy-like nanostructured organic film possesses a Raman EF of 3.4 $\times 10^3$ for MB as the probe molecule. Theoretical calculation and contrast experiment indicate that the π-conjugated core fluorocarbon substitution and the unique DFH-4T film morphology facilitate intermolecular CT between the organic semiconductor and the probe molecule. This finding provides a new material for design plasmon-free SERS substrates, which is the first time for π-conjugated organic small molecule film enhancing Raman signals in the plasmon-free SERS filed. Subsequently, *Demirel et al.* reported a nanostructured film of the small molecule DFP-4T, consisting of a fully π-conjugated diperfluorophenyl-substituted quaterthiophene structure (Fig. 9). Compared with 5,5‴-di(per-fluorophenylcarbonyl)-2,2′:5′,2″:5″,2‴-quaterthiophene (DFPCO-4T) and DFH-4T film, the DFP-4T film exhibited a remarkable SERS sensitivity with a high EF exceeding $10^5$ and a low LOD of $10^{-9}$ M for MB molecule [47]. According to the experimental and theoretical results, two conditions are required to satisfy for achieving high-performance SERS activity in organic semiconductors. One is that there exists small but nonzero oscillator strength in the CT state between the organic semiconductors and probe molecules, and the other is CT energy should be close to incident light to realize the resonance enhancement. It is expected to further improve the SERS activity of organic semiconductors by using chemical doping or electrical charge-injection which could tailor band edges and introduce free carrier into semiconductor structures. Simultaneously, as their structural flexibility and high tunability, organic semiconductors may achieve selective SERS detection toward specific analytes by adjusting their functional group and π-backbone length.

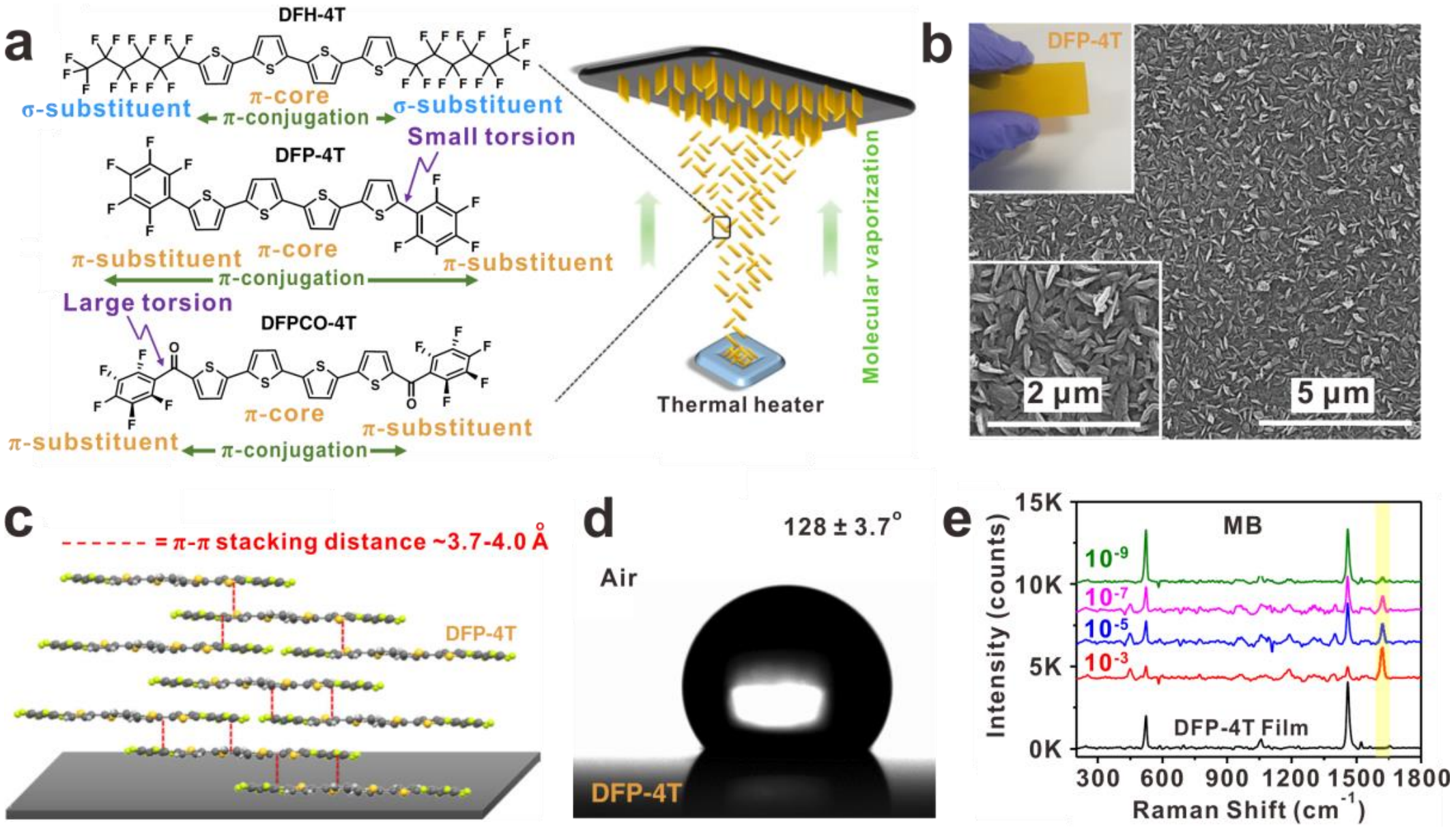

**Fig. 9** (a) Chemical structures of the 4T-based semiconductors and schematic illustration of the preparation process for nanostructured films, (b) SEM images, (c) molecular packing diagram in the out of plane direction, (d) contact-angle of DFP-4T films, and (e) The SERS spectra of MB probe on DFP-4T films at different concentrations. (Reprinted and adapted with permission from ref. [47]. Copyright 2019 Springer Nature.)

### 3.7 Other strategies

Additionally, other strategies that benefit to the performance of plasmon-free SERS substrates were proposed. As an example, *Miao et al.* reported the layer-number-dependent Raman enhancement effect on 1T′ phase $ReS_2$ [28]. It is found that the layer numbers from few-layer to monolayer can significantly strengthen the Raman enhancement effect on $ReS_2$ because the $ReS_2$ undergoes an indirect to direct band gap transition. As a result, monolayer $ReS_2$ exhibits ultrasensitive SERS activity and the LOD of R6G as low as $10^{-9}$ M. For monolayer $ReS_2$, the direct band gap allows excited electrons in CB to be immediately recombined with trapped hole carries, which is conducive to fluorescence quenching of dye molecules and CT process in molecule-$ReS_2$ system. However, in the terms of few-layer $ReS_2$, although the band gap shrinks slightly, the transition from the direct band gap to the indirect band gap may prolong the lifetime of excited electrons in the CB, resulting in a smaller difference in the chemical potential between probe molecules and $ReS_2$, which suppresses the CT process from the LUMO of molecules to the CB of $ReS_2$. *Hu et al.* demonstrated that electric field control of the SERS effect by adjusting the hole concentration of Sb:Na [139]. In the experiment, they dripped dye molecules onto the fabricated Sb:Na QDs-based field effect transistor (FET) device and observed the SERS effect *in situ*. More holes are injected into the FET device by applying different gate voltages, which enhances the SERS effect by boosting the CT between the molecules and Sb:Na QDs, bringing the LOD to the sub-femtomolar level. Similarly, *Zhou et al.* reported an electrically tunable SERS substrate based on $WO_{3-x}$ films [185]. The EF of which can increase from $3.01 \times 10^5$ to $1.14 \times 10^6$ by electrical programming of the defect density through the oxide leakage current control. Besides, highly tailorable MOFs have been used as SERS substrates with a low LOD of $10^{-8}$ M and a high EF of ~$10^5$, reported by *Zhao et al.* [42]. By adjusting the metal centers and organic ligands, the electronic band structures of the MOF-based SERS substrates could be purposively manipulated to match that of the target analyte, thus resulting in the combination of several resonances and subsequent prominent SERS enhancement. Further, by adjusting the LUMO and HOMO levels of ZIF-67 through a doping process with different metal ions, *Xu et al.* fabricated a flexible and reusable SERS substrate with a high EF ($6.07 \times 10^6$) [45]. *Li et al.* reported that surface-modified 2D $Ti_3C_2$ sheets, as SERS substrates, possess highly sensitive but nonselective enhancement with the detection limit down to the pM level [53]. The Raman signals of MB molecule adsorbed on $Ti_3C_2$-Al(OH)$_4$ sheets is four orders of magnitude higher than that of $Ti_3C_2$-OH/F sheets. The enhanced SERS activity could be attributed to the strong but nonselective interactions between substrate and analyte, which is induced by the surface aluminum oxyanion groups. With the aluminum oxyanion groups closely and evenly pack on the sheet surface, the analytes could adopt a configuration similar to that on a blank substrate, lying flat with C-N slightly distorted toward the substrate, which allows the unusual combination of ultrasensitive enhancement of signals and no obvious preference among different vibrational modes. *Lin et al.* proposed a low temperature-based strategy to facilitate the PICT processes and improve semiconductor SERS activity [130]. They found that the Raman intensity of 4-Mpy molecule adsorbed on porous ZnO nanosheets with abundant surface defect states is enhanced about 4 times at a 77 K compared to that at room temperature. Low-temperature condition could provide an ideal environment to reduce phonon-assisted relaxation and weaken lattice thermal vibration, thereby reducing the non-radiative recombination of excitons and increasing the number of photo-induced electrons to participate in the PICT processes. Finally, a low LOD of $1\times10^{-7}$ M and a high EF of $7.7\times10^{5}$ for 4-Mpy molecules are obtained at a low temperature of 77 K. *Zhao et al.* reported that high pressure could promote the CT efficiency and change the intensity of Raman signal in the semiconductor-molecule system [186, 187]. In addition, *Li et al.* investigated the effect of $MoS_2$ interlayer distances on the SERS activity of $MoS_2$ [188]. They found that $MoS_2$ with smaller interlayer distances shows stronger SERS enhancement and an EF as high as $5.31\times10^{5}$ could be achieved in $MoS_2$-0.62 (interlayer distance is 0.62 nm). In a word, these strategies not only provide a deep insight of the enhancement mechanisms for plasmon-free SERS materials, but also open up the way for preparation and application of plasmon-free SERS materials.

# 4 The applications of plasmon-free SERS

## 4.1 Biomedical diagnosis

Biomedical diagnosis with SERS is an effective diagnostic method because it is ultrasensitive, non-destructive, and capable of real-time molecular detection. Compared to noble metals, plasmon-free materials show superior biocompatibility and high stability [189, 190]. Furthermore, plasmon-free materials can form stable coordination bonds with the carboxyl and amine groups of biomolecules, maintaining their biological activity. Therefore, plasmon-free SERS is a promising method for biomedical diagnosis. *Haldavnekar et al.* obtained a ZnO-based semiconductor quantum SERS probe using femtosecond pulsed laser processing [129]. Owing to the surface oxygen vacancies and stacking faults, the probes could achieve the EF $\sim10^{6}$ with LOD down to $10^{-9}$ M. The ratio of peak intensities of lipids and proteins were used to distinguish between cancer and non-cancer. In this way, the probes achieve label-free, in vitro SERS diagnosis of cancer with a single-cell-level detection (Fig. 10). Apart from that, the organic semiconductor SERS sensor has been served for the detection of deoxyribonucleic acid (DNA) methylation and gene expression [191]. The reported organic semiconductor SERS sensor shows ultrasensitive SERS detectivity, which could detect the genomic DNA even at femtomolar concentration. We know that the genomic DNA can provide the information about the structural, molecular, and gene expression changes of DNA. In this case, this SERS sensor can be used for diagnosing the cancer stem cells by analyzing the differences between the genomic DNA of cancerous and non-cancerous cells. Besides, *Lin et al.* fabricated black $TiO_2$ nanoparticle that could enhance Raman scattering with the EF $4.3 \times 10^5$ [159]. On the one hand, the black $TiO_2$ nanoparticle could be applied to SERS mapping imaging technology, which has been employed to accurately diagnose the MCF-7 drug-resistant breast cancer cells. On the other hand, with excellent photothermal conversion efficiency, the black $TiO_2$ nanoparticle could also act as a promising heat mediator for efficient photothermal ablation of the cancer cells. Capable of detecting different gaseous indicators of lung cancer with ppm detection limit, the MIL-100(Fe) SERS platform proposed by *Fu et al.* shows great potential for screening and clinical diagnosis of lung cancer in the early stage [44]. In addition, *Tian et al.* reported that functionalized hexagonal boron nitride nanosheets could be used as a SERS platform for real-time monitoring and imaging of microRNA [39]. All these works indicate that plasmon-free SERS substrates have a bright prospect in cell labeling and imaging, biological diagnosis, photothermal therapy and so on.

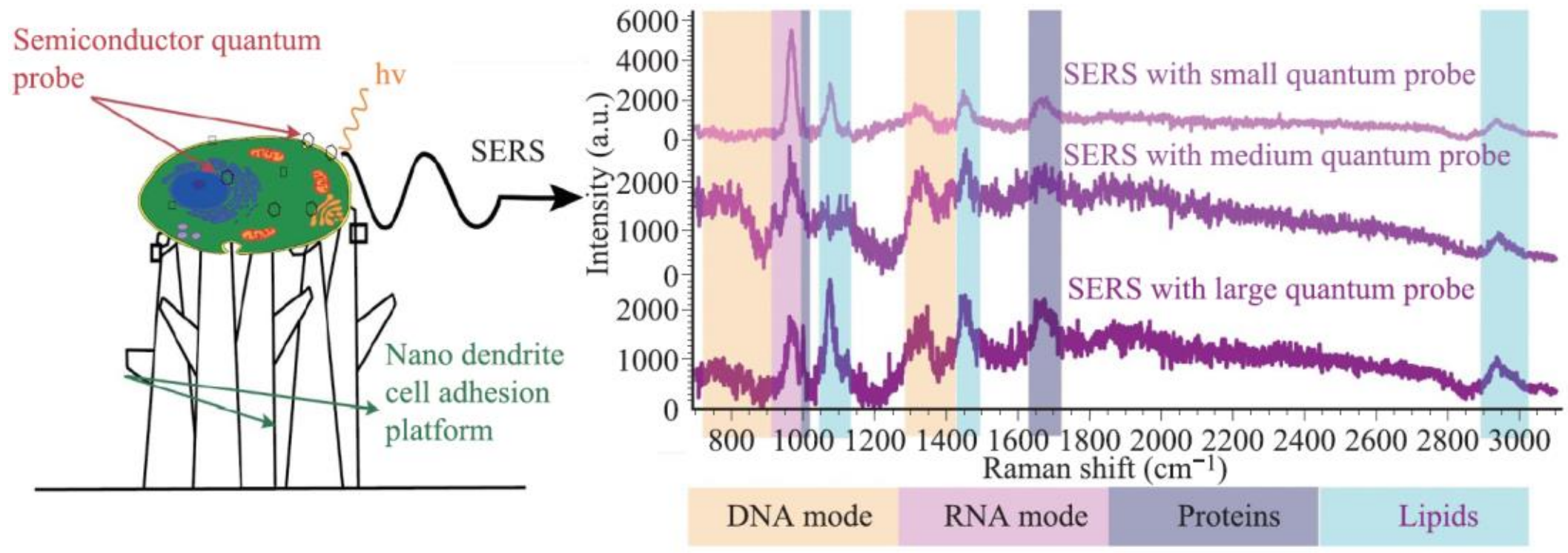

**Fig. 10** Different sizes of plasmon-free semiconductor quantum SERS probes as the platforms for in vitro cell detection. (Reprinted and adapted with permission from ref. [129]. Copyright 2018 Springer Nature.)

### 4.2 Metal ions and organic pollutants sensing

Metal ions and organic pollutants sensing have aroused wide concern for their strong environmental and biomedical impact [192]. Fortunately, the selectivity and reusability of plasmon-free SERS substrates provide excellent platforms for metal ions and organic pollutants sensing. One case is the "turn-off" SERS strategy for metal ion detection proposed by *Ji et al.* [193]. Wherein the stable and highly reproducible SERS intensity of alizarin red S (ARS)-$TiO_2$ complexes was borrowed from the efficient CT process that results from the vibronic coupling between the excited molecular level and the CB of $TiO_2$. They found that the SERS intensities of the complexes are sensitive to the Cr(VI) concentration due to co-catalysis, and the LOD of 0.5 μM Cr(VI) cations has been achieved. Furthermore, the complexes exhibit a remarkably high selectivity toward Cr(VI) cations, which originates from the energy level matching between the redox potential of Cr(VI)/Cr(V) and the energy level of the $TiO_2$. Our group also prepared a series of transition metal oxide chips by a general strategy based on magnetron sputtering with $H_2$ annealing treatment. Successfully, these chips could selectively improve the signals of specific organic pollutant in the detected mixture solutions [105]. In the terms of practical application, plasmon-free materials could be used as recyclable SERS substrates owing to their self-cleaning ability by photo-degrading the organic pollutants. *Majee et al.* applied few-layer $MoS_2$ as a SERS substrate for the ultrasensitive detection of organic compounds at nanomolar level [194]. And, the treatment of adsorbed organic compounds over few-layer $MoS_2$ under visible light illumination has also been demonstrated to verify the reusability of the SERS substrate with repeated detectability.

### 4.3 Chemical and biochemical reactions monitoring

Typically, as a powerful approach for monitoring chemical and biochemical reactions, plasmon-free SERS has been used for studying surface/interface redox reactions, molecular adsorption, electrochemical reactions kinetics, and so on [195]. For example, with the help of electrochemical anodization of titanium, *Han et al.* prepared nanostructured $TiO_2$ electrodes and altered the Raman EF of cyt $b_5$ adsorbed on $TiO_2$ electrodes by adjusting the surface roughness with different anode voltages [196]. According to the SERS spectra measured at different electrode potentials, the protein oxidation marker band ν4 located in 1374 $cm^{-1}$ (oxidized state) shifted to 1360 $cm^{-1}$ (reduced state), indicating a redox protein process has happened. *Alessandri et al.* demonstrated that $SiO_2$-$TiO_2$ core-shell colloidal crystals can be used as a SERS probe to monitor the glutathione (L-γ-glutamyl-cysteinyl-glicine, GSH) redox cycle at physiological concentration in aqueous environment [197]. When GSH is oxidized to its disulphide derivative, the signal of -SH stretching would disappear in SERS spectrum. Thus, GSH oxidation process can be monitored by observing the -SH stretching Raman signal. *Glass et al.* reported that semiconductor-based SERS enhancements could be used as a direct method of surface oxygen vacancies concentration quantification [198]. Besides, our group reported that the oxidation process of ultraviolet-ozone-treated $MoS_2$ could be successfully monitored by using plasmon-free SERS technology [134]. During the oxidation process, the changes in energy bands caused by defects in $MoS_2$ would remarkably affect SERS. Thus, we could quantify the degree of oxidation at different treatment time according to the exponential relationship between the percent of oxidation and treatment time obtained by SERS. Experimental and theoretical results indicate that $MoS_2$ is first doped with $O_{ads}$, then doped with $O_s$, and finally there forms a completely oxidized layer ($MoO_3$).

Our study may provide a general strategy for tracking the oxidation and degradation process of other 2D materials.

### 4.4 Photoelectric characterization

Since that the intrinsic properties of plasmon-free SERS materials, such as the band structure and carrier concentration, affect the SERS property and other photoelectric characteristics of the materials, SERS measurement could be served as a tool to indirectly explore the photoelectric performance. Based on this point, *Chen et al.* monitored the carrier dynamics of π-conjugated polymer using SERS technology [199]. With poly(3,4-ethylenedioxythiophene):poly(styrenesulfonate) (PEDOT:PSS) as the plasmon-free SERS substrate, they discussed the Raman characterizations during the carrier dynamics process under different voltages and conductive situations. Due to abundant delocalized π electrons in PEDOT:PSS, the induced chemical enhancement causes high EF of the material. Furthermore, different bias voltages increased or reduced the carrier density of the PEDOT:PSS substrate, leading to the blue shift or red shift in Raman peaks respectively. This discovery provides an important theoretical fundament for the chemical enhancement model in SERS and offers a new idea for the researches on photoelectric devices. Similarly, *Yu et al.* constructed a photodetector device based on 4-Mpy-modified perovskite. The results demonstrated that a correlation between the CT resonance-enhanced Raman and the photoelectronic responses of perovskite materials [31]. By exploring the CT pathways of hybrid dye/molecule-metal oxide complexes via SERS, *Tarakeshwar et al.* discovered the close correlation between the Raman enhancement and the electron transfer rate on the molecule-$TiO_2$ interface, providing a significant guidance for the nature of CT paths [200]. Besides, *Yang et al.* contrasted the photoluminescence and SERS spectra of the molecules absorbed on ZnO with that on Nd-doped ZnO, proving that unique property of $Nd^{3+}$ ions not only can improve the SERS signals but also can eliminate the SERS fluorescence background [124]. They put forward a new path for studying CT and explaining the fluorescence quenching. Coincidentally, *Tao* and his colleagues studied the fluorescence quenching of plasmon-free SERS materials and drew a conclusion that it is helpful for the Raman enhancement [25].

## 5 Summary and outlook

Plasmon-free substrates represent new frontier for SERS. There has been surprising and accelerated development of plasmon-free SERS over the past decade. In this review, we introduce the background, enhancement mechanisms, enhancement strategies, and applications of ultrasensitive SERS sensing beyond plasmonics. The background of SERS and the advantages of plasmon-free SERS are firstly stated, followed by the enhancement mechanisms of plasmon-free SERS. Then, we summarize numerous enhancement strategies for constructing ultrasensitive plasmon-free SERS substrates. After that, the promising application of plasmon-free SERS in various areas such as biomedical diagnosis, metal ions and organic pollutants sensing, chemical and biochemical reactions monitoring, and photoelectric characterization are summarized. Considering recent development on plasmon-free SERS, we propose some potential significant topics in the field: (1) developing new technologies to prepare the plasmon-free SERS substrates. Some technologies like solvothermal method [16], CVD [25], magnetron sputtering [64], electrochemical method [196] have been developed to obtain plasmon-free SERS substrates. However, these preparation processes are generally complicated, low-yield and expensive. It is necessary to fabricate ultrasensitive plasmon-free SERS substrates through a simple, convenient and large-scale approach, which is the basis of their wide applications. (2) Exploring new strategies to improve the SERS activity of plasmon-free substrates. At present, plasmon-free SERS materials have shown high sensitivity by using different enhancement strategies, but there is still a big gap with noble metal materials whose EFs are even higher than ~$10^{14}$ for single-molecule detection [201]. High-performance SERS substrates are still the key factor limiting the practical application of plasmon-free SERS. In addition, researchers recently reported that some noble metal-free materials such as metal oxides [202-209], metal carbides [9, 210] and metal nitrides [211-213] possess high charge carrier concentration, which could induce LSPR effect in the visible region [214]. Plasmon resonance together with CT resonance, exciton resonance, molecular resonance, and Mie resonance may further improve the SERS activity of noble metal-free materials under some specific conditions. (3) Discovering novel plasmon-free SERS materials. In recent years, a series of novel plasmon-free SERS materials have been explored (TMDs, MOFs, MXenes, perovskites, organic semiconductors and so on) [25, 26, 29, 31, 32, 42, 43, 46, 47]. New materials not only can expand people's cognition of plasmon-free SERS materials, but also benefit further understanding of the enhancement mechanism. (4) Expanding the application of plasmon-free SERS substrates. Plasmon-free SERS substrates with ultrahigh sensitivity have been reported in recent, showing the great potential for replacing plasmonic noble metal substrates. It is worth noting that the high biocompatibility and low biological damage of plasmon-free SERS substrates have given them an inherent advantage in biochemical and biomedical sensing. Plasmon-free SERS can also be integrated into other techniques (IR spectroscopy, fluorescence spectroscopy, atomic force microscope, and so on) [215-217], which can achieve multiplexed analysis. Moreover, it may be a promising application direction of the plasmon-free SERS materials to combine novel plasmon-free materials with conventional noble metals [218-223]. This integration not only could enable a synergistic modulation of both the chemical and electromagnetic properties of the hybrid SERS platforms and obtain an ultrasensitive SERS activity, but also could take the advantages of both materials to achieve multifunctional applications.

## Acknowledgements

This work was supported by the National Natural Science Foundation of China (Grant No. 11874108) and the National Key R&D Program of China (Grant No. 2017YFA0403600).